\documentclass[aps,twocolumn,amsmath,amssymb,reprint,floatfix,superscriptaddress, longbibliography,prx]{revtex4-2}
\usepackage{amsmath,amssymb,bm,graphicx}
\usepackage{times}
\usepackage{xcolor}
\usepackage{enumitem}
\usepackage{overpic}
\usepackage{float}

\usepackage{hyperref}
\hypersetup{
colorlinks=true,
linkcolor=blue,           
citecolor=blue,           
filecolor=magenta,        
urlcolor=cyan
}
\allowdisplaybreaks

\newcommand{\bra}[1]{\langle #1|}
\newcommand{\ket}[1]{| #1 \rangle}
\newcommand{\average}[1]{\langle #1 \rangle}

\newcommand{\papertitle}{A Quantum Mechanical Pendulum Clock}

\begin{document}

\title{\papertitle}

 \author{Matteo~Brunelli}
 \thanks{These two authors contributed equally}
\affiliation{Department of Physics and Swiss Nanoscience Institute, University of Basel, Klingelbergstrasse 82, 4056 Basel, Switzerland}
\affiliation{JEIP, UAR 3573 CNRS, Coll\`ege de France, PSL Research University, 11 Place
Marcelin Berthelot, 75321 Paris Cedex 05, France}
\author{Mohammad Mehboudi}
\thanks{These two authors contributed equally}
\affiliation{Vienna Center for Quantum Science and Technology, Atominstitut, TU Wien, Vienna 1020, Austria}

\author{Nicolas Brunner}
\affiliation{Département de Physique Appliquée, Université de Genève, 1211 Genève, Switzerland}

\author{Patrick P. Potts}
\affiliation{Department of Physics and Swiss Nanoscience Institute, University of Basel, Klingelbergstrasse 82, 4056 Basel, Switzerland}


\begin{abstract}
We investigate an optomechanical system as a model of an autonomous mechanical pendulum clock in the quantum regime, whose operation relies only on incoherent (thermal) resources. The escapement of the clock, the mechanism that translates oscillatory motion into ticks, is provided by an emitter in the optical cavity and the operation of the clock relies on the existence of a limit cycle. Since the clock is based on an oscillatory
degree of freedom, it can overcome the thermodynamic uncertainty relation and is thus more accurate than clocks that rely only on stochastic transitions. Furthermore, by increasing the amount of emitters in the cavity, the clock approaches the behavior expected for a macroscopic pendulum clock, where fluctuations become irrelevant while the clock dynamics becomes completely irreversible. This allows for investigating the quantum-to-classical transition of pendulum clocks.
\end{abstract}

\maketitle

\section{Introduction}
The nature of time has puzzled and fascinated humanity since millennia. A pragmatic answer to the question ``what is time?" is provided by the famous statement: ``Time is what a clock measures", attributed to Einstein. This raises the question of the necessary ingredients for a system to constitute a clock. 
Any system that operates as a clock requires irreversibility: the heat radiated by the sun in sundials, the lowering weights in a pendulum clock, or the dissipated laser power in an atomic clock. In recent years, thermodynamic principles have been employed to quantify the amount of dissipation that is needed for a clock to operate with a certain accuracy~\cite{Barato_2016_brownian,erker:2017,Milburn_2020_thermodynamicsclocks,pearson:2021,schwarzhans:2021,Pietzonka_2022_pendulum,meier:2023,He_2023_measurementclock,Gopal_2024_electronicclock,culhane:2024,Scheer_2024_josephsonclock,prech:2024, meier:2024,Wadhia:2025}.

A particularly prominent constraint is provided by the thermodynamic uncertainty relation (TUR)~\cite{Barato_2015_tur,gingrich:2016,Horowitz2020}, which states that the accuracy~\footnote{We note that we use the accuracy as defined in Refs.~\cite{erker:2017,meier:2023}, which can be understood as the amount of ticks until the clock is off by one tick on average. A large accuracy is thus desired. This is to be contrasted with atomic clocks, where the accuracy is typically quantified in terms uncertainties which are desired to be small.} is upper bounded by the entropy production, a common measure of irreversibility.
The TUR provides a useful benchmark and a linear scaling of accuracy with entropy production has indeed been found both theoretically~\cite{Barato_2016_brownian,erker:2017} as well as experimentally~\cite{pearson:2021} in different implementations. However, the TUR only holds for classical systems that obey Markovian rate equations or overdamped dynamics. This is not the regime where most clocks operate. Instead, most clocks that are used for timekeeping rely on oscillatory degrees of freedom, which are not constrained by the TUR. Still, the TUR remains a relevant benchmark to characterize their performance (for other constraints, see Refs.~\cite{woods:2022,woods:2023,meier:2023,Scheer_2024_josephsonclock}). Violations of the TUR have been found in classical pendulum clocks~\cite{Pietzonka_2022_pendulum,Gopal_2024_electronicclock}, as well as in systems that rely on quantum coherence~\cite{woods:2023,culhane:2024,meier:2024}. Nevertheless, the limitations and required resources of oscillating clocks have only recently started to attract attention~\cite{Pietzonka_2022_pendulum,Gopal_2024_electronicclock,culhane:2024,Scheer_2024_josephsonclock}. 

Here, we address this gap by investigating a quantum-mechanical pendulum clock based on an optomechanical system, where the mechanical oscillator provides the pendulum.
Different from previous models of autonomous quantum clocks~\cite{erker:2017, schwarzhans:2021, WoodsHorodecki2023, Manikandan2023, woods:2022}, our model introduces a quantum mechanical version of the ingenious escapement mechanism found in classical pendulum clocks. The escapement consists of a single emitter with three energy levels coupled to a cavity field, and
relies on the mechanism of Rabi oscillations to affect the mechanical motion. It implements the smallest possible quantum escapement, in a similar spirit to other minimal models of quantum thermal machines~\cite{Linden2010, Gelbwaser-Klimovsky2013, Venturelli2013}.
 Our model is only driven by time-independent thermal resources and is thus autonomous, akin to the time-independent force of the weights in a classical pendulum clock. This allows for identifying all necessary resources for running the clock. As expected, we find that the accuracy of the clock monotonously increases with entropy production, but the TUR is violated. Upon increasing the number of emitters, our clock behaves more and more like a classical pendulum clock, as fluctuations are suppressed and both the accuracy and the entropy production increase, demonstrating the irreversible nature of accurate time keeping. 

Our results show that a realistic quantum mechanical model can describe a pendulum clock from the quantum to the classical regime. Pendulum clocks can thus be implemented in the quantum regime, with accuracies that overcome the bound provided by the TUR. Our work thereby opens a new avenue for investigating the limitations and resources of oscillating clocks, providing insight into the physics of timekeeping and noise engineering, and potentially resulting in more efficient time-keeping devices for quantum technology.

This paper is structured as follows: In Sec.~\ref{sec:model}, we introduce the model for the quantum-mechanical pendulum clock and we identify the constituents found in a conventional pendulum clock. Section \ref{sec:operation} illustrates that the model indeed operates as a clock, with periodic motion of the pendulum and ticks that occur twice per mechanical period. In Sec.~\ref{sec:clockperf} we quantify the performance of the clock and a thermodynamic analysis follows in Sec.~\ref{sec:thermodynamics}. Section \ref{sec:classical_limit} investigates  multiple emitters and we conclude in Sec.~\ref{sec:conclusions}.

\begin{figure*}[t]
\centering
\includegraphics[width=1.\textwidth]{ 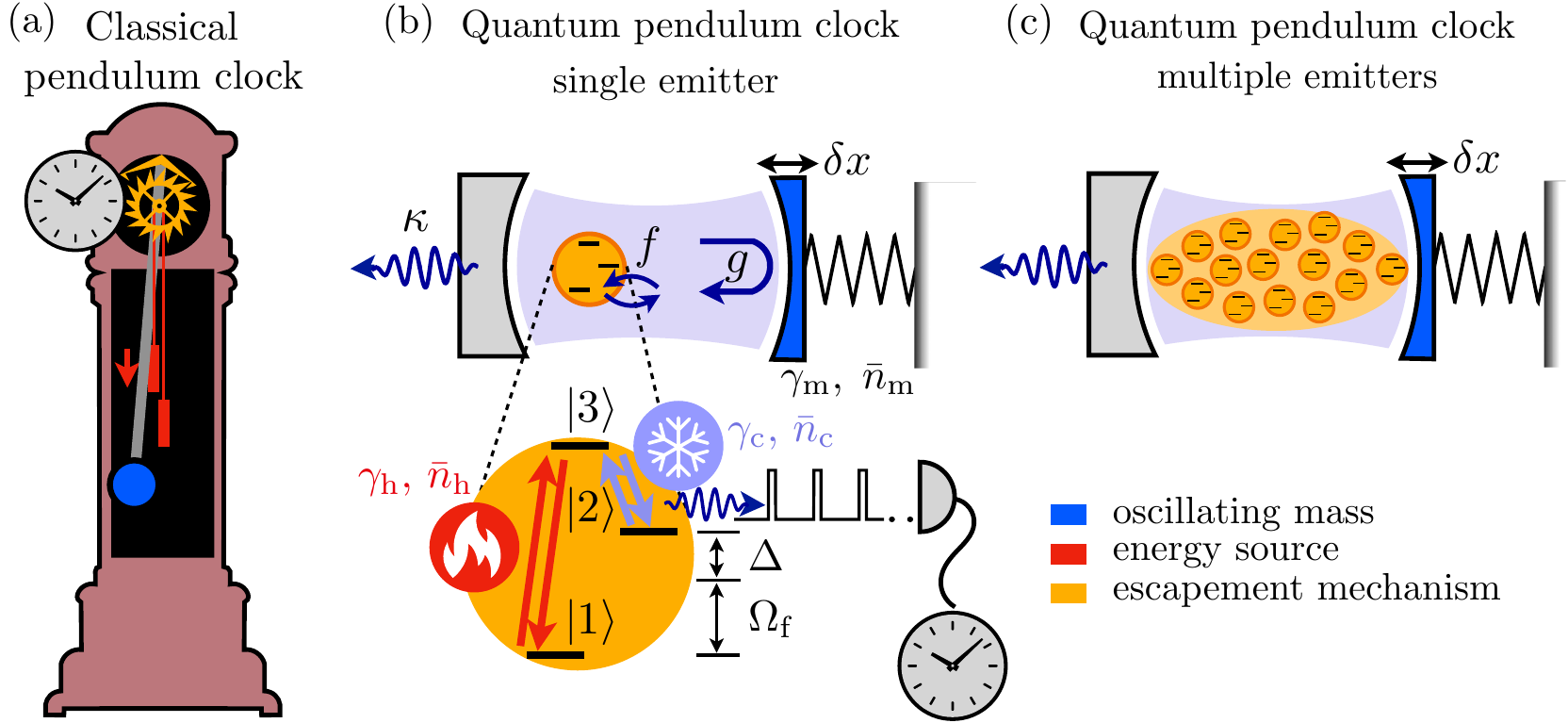}
\caption{\label{f:Sketch} 
Classical and quantum mechanical pendulum clock.
(a) Grandfather clock as an example of classical pendulum clock. A swinging pendulum (blue) subject to unavoidable friction maintains its periodic motion by utilizing the energy stored in suspended weights (red). The energy is transferred through the escapement mechanism (yellow), which imparts small impulses to the pendulum at regular intervals. This mechanism converts the gravitational energy of the descending weights into discrete impulses (tick-tock) that sustain the pendulum's oscillation and move the clock's hands. 
(b) Quantum version of a pendulum clock implemented in an optomechanical system. A mechanical resonator (blue) is in contact with a finite temperature bath that provides dissipation and fluctuations. The mechanical displacement  is coupled via radiation-pressure to a cavity, which is near resonant to the lasing transition ($\ket{1}\leftrightarrow \ket{2}$) of a three level emitter (yellow). The emitter acts as escapement, timely releasing/injecting photons into the cavity. Photons are loaded into the cavity by a hot bath (red). The ticks of the clock are provided by the photons emitted into the cold bath. (c) Many-emitter version of the quantum pendulum clock, in which many identical emitters are coupled to the cavity field. This configuration recovers the classical limit of an ideal (noiseless) pendulum clock.}
\end{figure*}

\section{The optomechanical pendulum clock}
\label{sec:model}

Pendulum clocks, like the grandfather clock in Fig.~\ref{f:Sketch}\,(a), are a testament to human ingenuity and provided the most accurate timekeeping devices for centuries~\cite{encyclopedia}. Their general working principle consists of counting the periods of a swinging pendulum by means of an escapement mechanism. The escapement mechanism moves a pointer (the hands of the clock) in the forward direction by discrete amounts. At the same time, it transfers energy from a descending weight to the pendulum to compensate for the oscillator's damping while minimally affecting its frequency. This ingenious mechanism revolutionized timekeeping devices, allowing for a dramatic boost in accuracy.

A mechanical pendulum clock consists of the following key elements, shown in Fig.~\ref{f:Sketch}\,(a):
i) A pendulum that serves as a timekeeping element, swinging at a consistent frequency; the gravitational force acts as the restoring force.
ii) An energy source that keeps the pendulum swinging, compensating the  energy  dissipated to friction; the descending weights store potential energy that powers the clock's motion.
iii) An escapement mechanism to transfer energy from the weights to the pendulum, providing periodic impulses to maintain its motion. The escapement mechanism also serves a second purpose. It regulates the pendulum motion converting the continuous oscillation of the pendulum into discrete steps---the ``tick-tock" sound---which, through a series of gears, ultimately move the clock's hands.  It takes a ``tick'' and a ``tock" for the pendulum to complete one full swing. These three elements are highlighted in Fig.~\ref{f:Sketch}\,(a) in blue, red and yellow, respectively.

We propose a self-contained quantum mechanical model of a pendulum clock. We first describe the model and then illustrate the counterparts of the elements i)-iii) that allows it to operate as a quantum pendulum clock.   
The system that we consider, shown in Fig.~\ref{f:Sketch}\,(b), is an  optomechanical system coupled to a three-level emitter. It consists of a mechanical (optical) resonator with frequency $\Omega_\mathrm{m}$ ($\Omega_\mathrm{f}$), described by annihilation operator $\hat b$ ($\hat a$) and a three-level system with levels $\ket{j}$, $j=1,2,3$, and corresponding energies $\varepsilon_1<\varepsilon_2<\varepsilon_3$. The free Hamiltonian of the system is given by ($\hbar=1$)
\begin{equation}
\label{eq:hamfree}
   \hat H_0=\sum_{j=1,2,3}\varepsilon_j\hat{p}_j + \Omega_\mathrm{f}\hat a^\dagger \hat a + \Omega_\mathrm{m}\hat b^\dagger \hat b \,,
\end{equation}
with $\hat p_{j}=|j\rangle\langle j|$. In the following, we will also indicate the detuning between the cavity and the $\ket{1}\leftrightarrow \ket{2}$ transition by $\Delta=\varepsilon_2-\varepsilon_1-\Omega_\mathrm{f}$ and the transition operators of the emitter by $\hat\sigma_{ij}=\ket{i}\bra{j}$.

The electromagnetic field inside the cavity is coupled to both the mechanical resonator and the emitter. It is coupled to the $\ket{1}\leftrightarrow \ket{2}$ transition with strength $f$ and is radiation-pressure coupled to the mechanical displacement $\hat x_\mathrm{m}= (\hat b+\hat b^\dagger)/\sqrt2$, with single-photon optomechanical coupling $g$. In Fig.~\ref{f:Sketch}\,(b) we illustrate the case of a Fabry Perot cavity with a movable end mirror but several other designs are possible~\cite{Aspelmeyer_2014_review}.
The total Hamiltonian of the system is given by  
\begin{equation}
\label{eq:ham}
\hat H= \hat H_0 + f (\hat a^\dagger \hat{\sigma} + \hat a \hat{\sigma}^\dagger) -\sqrt{2}g \hat a^\dagger \hat a \hat{x}_\mathrm{m}\,,
\end{equation}
where for simplicity we set $\hat \sigma_{12}=\hat \sigma_{21}^\dagger\equiv \hat \sigma$. The sign of the optomechanical coupling is due to the fact that we take positive values of the displacement to indicate an
increase in cavity length, and therefore a decrease of the frequency. We notice that the total Hamiltonian preserves the number of emitter plus cavity excitations $\hat N=\hat a^\dagger \hat a + \tfrac12 \hat \sigma_z$, with $\hat \sigma_z=\ket{2}\bra{2}-\ket{1}\bra{1}$.

Besides the Hamiltonian terms, we also need to model the dissipative processes, see Fig.~\ref{f:Sketch}\,(b).  
The mechanical (optical) resonator is in contact with a thermal bath at finite temperature with occupation $\bar n_\mathrm{m}=(e^{\beta_\mathrm{m}\Omega_{\rm m}}-1)^{-1}$ [$\bar n_\mathrm{f}=(e^{\beta_\mathrm{f}\Omega_{\rm f}}-1)^{-1}]$ and rate $\gamma_\mathrm{m}$ ($\kappa$). The emitter is coupled to two bosonic reservoirs: the transition $\omega_{13}=\varepsilon_3-\varepsilon_1$ is coupled to a hot bath with occupation $\nobreak{\bar n_\mathrm{h}=(e^{\beta_\mathrm{h}\omega_{13}}-1)^{-1}}$ at a rate $\gamma_\mathrm{h}$, while the transition $\omega_{23}=\varepsilon_3-\varepsilon_2$ is coupled  to a cold bath with occupation $\bar n_\mathrm{c}=(e^{\beta_\mathrm{c}\omega_{23}}-1)^{-1}$ at a rate $\gamma_\mathrm{c}$, where $\beta_\mathrm{\alpha}=1/T_{\alpha}$ is the inverse temperatures of bath $\alpha$---we set $\hbar=k_{\rm B}=1$ throughout the paper. 
We model the dissipative processes described above via the following master equation
\begin{equation}\label{eq:FullME}
\frac{\mathrm{d}\hat \varrho_{\rm tot}}{\mathrm{d}t}=-i[\hat H, \hat \varrho_{\rm tot}] + \mathcal{L}_\mathrm{f}\hat \varrho_{\rm tot} +  \mathcal{L}_\mathrm{h}\hat \varrho_{\rm tot} + \mathcal{L}_\mathrm{c}\hat \varrho_{\rm tot} + \mathcal{L}_\mathrm{m}\hat \varrho_{\rm tot}, 
\end{equation}
which includes, in order, the thermal bath coupled to the cavity, the two thermally-driven emitter transitions and the finite temperature mechanical phonon bath, where the thermal dissipators are given by 
\begin{align}
\mathcal{L}_\mathrm{f}\hat \varrho&=\kappa (\bar n_\mathrm{f}+1)\mathcal{D}[\hat{a}]\hat \varrho+\kappa \bar n_\mathrm{f}\mathcal{D}[\hat{a}^\dagger]\hat \varrho \,,\\
\mathcal{L}_\mathrm{h}\hat \varrho&=\gamma_\mathrm{h} (\bar n_\mathrm{h}+1)\mathcal{D}[\hat{\sigma}_{13}]\hat \varrho+\gamma_\mathrm{h} \bar n_\mathrm{h}\mathcal{D}[\hat{\sigma}_{31}]\hat \varrho \,,\\
\mathcal{L}_\mathrm{c}\hat \varrho&=\gamma_\mathrm{c} (\bar n_\mathrm{c}+1)\mathcal{D}[\hat{\sigma}_{23}]\hat \varrho+\gamma_\mathrm{c} \bar n_\mathrm{c}\mathcal{D}[\hat{\sigma}_{32}]\hat \varrho \,,\\
\mathcal{L}_\mathrm{m}\hat \varrho &=\gamma_\mathrm{m} (\bar n_\mathrm{m}+1)\mathcal{D}[\hat b]\hat \varrho+\gamma_\mathrm{m} \bar n_\mathrm{m} \mathcal{D}[\hat b^\dagger]\hat \varrho \,,
\end{align}
with $\mathcal{D}[\hat O]\varrho=\hat O \hat \varrho \hat O^\dagger -\frac12 \hat O^\dagger \hat O  \hat \varrho -\frac12 \hat \varrho \hat O^\dagger \hat O$.
We refer to the model in Eq.~\eqref{eq:FullME} as a quantum mechanical pendulum clock.

Let us now address how the elements i)-iii) are implemented in our model. The mechanical resonator clearly plays the role of the oscillating mass i).
The hot bath can then be seen as the energy source  
driving the clock ii). The damping of mechanical oscillations is counteracted by momentum transfer from the photons, 
which populate the cavity thanks to the presence of a hot and a cold bath driving the emitter transitions.  
Finally, the three-level emitter plays  the role of the escapement mechanism iii). As we will see, when on resonance the emitter releases timely photon pulses in the cavity, that keeps the clock running. These in turn are accompanied by photon emissions into the cold bath. Resolving these emission events yields the clock's ticking times. 

An essential aspect of our model concerns the way in which the system is driven. Indeed, Eq.~\eqref{eq:ham} lacks any coherent drive. This is because a self-contained description of the clock cannot resort to a coherent source of radiation, which represents the standard scenario for laser-driven optomechanical cavities~\cite{Aspelmeyer_2014_review}. In fact, a laser drive would provide an external well defined frequency reference, which effectively constitutes a clock of its own. Instead, we achieve the autonomous operation of the clock by coupling the cavity to a single, thermally driven emitter, which realizes the simplest possible lasing medium. 

To better appreciate this point we consider the non-interacting limit $f=g=0$, in which the emitter reaches a steady state characterized by $\langle \hat \sigma_z\rangle_\mathrm{ss}=(\bar n_\mathrm{h}-\bar  n_\mathrm{c})/(\bar n_\mathrm{c}+\bar n_\mathrm{h}+3\bar n_\mathrm{h}\bar n_\mathrm{c})$, see App.~\ref{app:three-level-laser} for details. From this expression we conclude that population inversion $\langle \hat \sigma_z\rangle_\mathrm{ss}>0$ occurs as long as $\bar n_\mathrm{h}>\bar n_\mathrm{c}$. 
In the presence of emitter-light coupling $f\neq 0$, population in state $\ket{2}$ is converted into cavity photons and lasing can be sustained. In this case, our system corresponds to a thermally driven three-level maser~\cite{ScovilSchulz-DuBois_1959_maser, Niedenzu_2019_work, Li_2017_maser}, which achieves lasing by only relying on the population imbalance between two emitter transitions.
For quantitative details we refer to App.~\ref{app:three-level-laser}.

\begin{figure*}[t]
\centering
\includegraphics[width=0.99\linewidth]{ 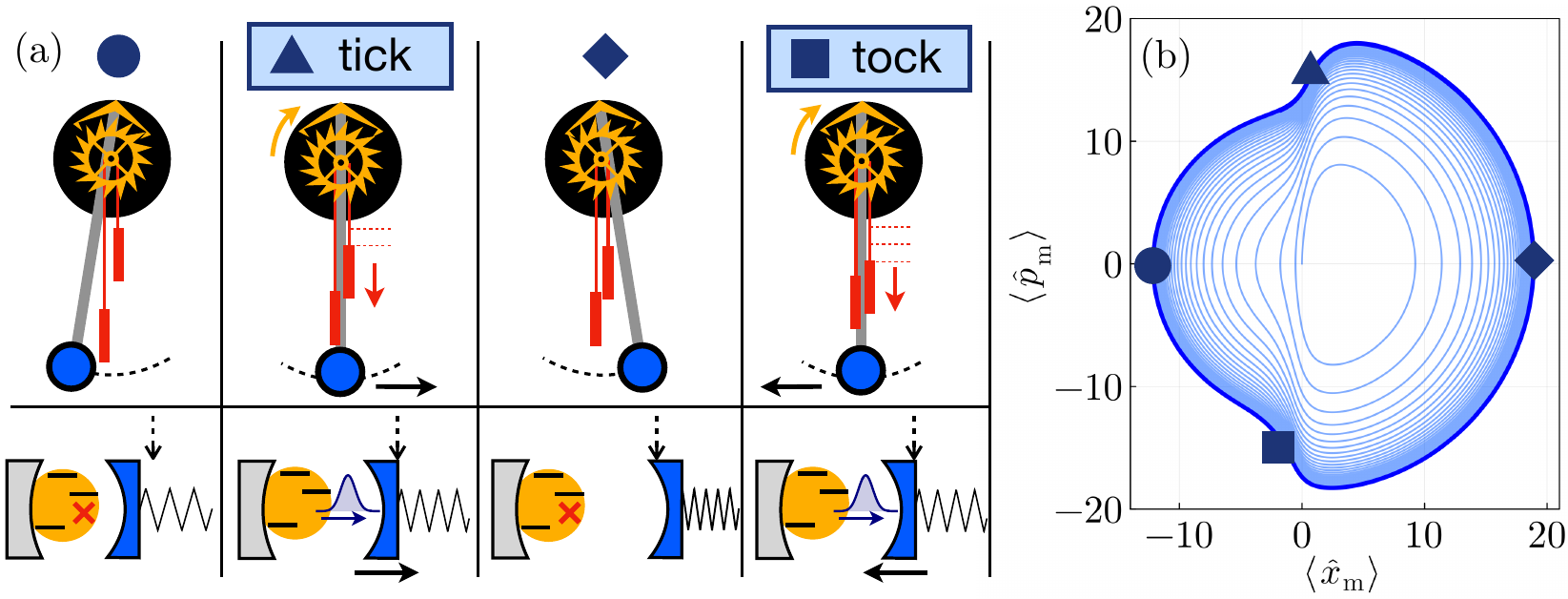}
\caption{\label{f:LimitCycle} Clock operation. (a) The pendulum clock operates in a cycle to complete a mechanical period. The four steps of the cycle are highlighted in the top row, see text for explanation, with their counterpart in our model the bottom row. The dashed black arrows in the bottom row indicate the mechanical equilibrium position.
(b) Exemplary mechanical  trajectory in phase space, showing the evolution of the expectation value of the mechanical position and momentum starting from the origin (light blue) and settling into a limit cycle (solid blue). Parameters are: $\omega_{13} = 240\gamma_{\rm h}, \omega_{23} = \Omega_{\rm f} = 120\gamma_{\rm h}$ ($\Delta=0$),}
$\Omega_\mathrm{m}=0.9\gamma_{\mathrm h}$, 
$f=20 \gamma_{\mathrm h}$,
$g=30\gamma_{\mathrm h}$,
$\kappa=10\gamma_{\mathrm h}$,
$\gamma_\mathrm{c}=100\gamma_{\mathrm h}$, $\gamma_\mathrm{m}=0.01 \gamma_\mathrm{h}$,
$\bar n_\mathrm{h}=10$, and 
$\bar n_\mathrm{c}=\bar n_\mathrm{m}=\bar n_{\mathrm f}=0$. 
\end{figure*}

\section{Clock operation}
\label{sec:operation}

After having identified the clock's constituents i)-iii) in our model, it remains to be determined whether the system~\eqref{eq:FullME} can operate as a clock.
We first recall the working cycle of a standard pendulum clock, exemplified by the grandfather clock of Fig.~\ref{f:Sketch}\,(a) and then discuss how an analogous situation can be realized in our quantum system. 
The operation cycle of a pendulum clock consists of four steps, depicted in Fig.~\ref{f:LimitCycle}\,(a):  
\begin{enumerate}[itemsep=1pt, left=0pt]
\item The cycle starts from the point of maximum excursion, where the pendulum is at rest and the escapement is locked.
\item The pendulum swings under the action of the gravitational force, reaching its fastest speed at the center. After passing the center, it releases a tooth of the escapement wheel. The release produces the ``tick" sound and allows the escapement wheel to rotate, pulled by the  weight. The escapement imparts a small push to the pendulum, compensating for the energy dissipated to friction.
\item The pendulum continues the swing until it reaches the turning point on the opposite side.
\item The pendulum swings back through the center in the opposite direction. Past the center, it releases the next tooth of the escapement wheel, producing the ``tock" and imparting a small kick to the pendulum. The cycle repeats.
\end{enumerate}	

We notice that in a real pendulum clock, the ticks and tocks happen when the escapement mechanism engages, which occurs slightly after the pendulum passes through the center. We will recover this behaviour also in our model, see Fig.~\ref{f:LimitCycle}\,(b). For the sake of simplicity, in the sketch of Fig.~\ref{f:LimitCycle}\,(a) we ignore this subtlety and put the ticks and tocks in correspondence with the pendulum passing through the centre.

\subsection{Semi-classical analysis}
\label{sec:semi-classical}

Our quantum mechanical model can operate in a cycle similar to the pendulum clock cycle described above, as shown in Fig.~\ref{f:LimitCycle}\,(b). 
A prerequisite for a working pendulum clock is to sustain periodic motion despite the presence of dissipation. 
In our model, the mechanical resonator is driven by the photons injected into the cavity, compensating for the energy loss due to friction. To build some physical intuition, consider the lasing transition to be resonant with the cavity, i.e., $\Delta=0$. Ideally, for a strong enough optomechanical coupling, a single cavity photon can displace the mechanical oscillator enough to bring the cavity out of resonance such that the injection of a second photon is suppressed. Once the photon leaks out of the cavity, the mechanical restoring force brings the cavity back to resonance, favoring the injection of the next photon into the cavity. 
This autonomous self-regulating mechanism can attain self-sustained mechanical oscillations, as we now show.

Since the Hilbert space dimension required to simulate the mechanical oscillations with the desired amplitude is prohibitively big, we resort to a mean-field treatment of the mechanical oscillator. To this end, we model the emitter and cavity through a master equation that depends on the average position of the mechanical oscillator. To this end, we replace the position operator in Eq.~\eqref{eq:ham} by its average, resulting in the master equation
\begin{equation}\label{eq:MFME}
\frac{\mathrm{d}\hat \varrho}{\mathrm{d}t}=-i[\hat H(\langle \hat{x}_{\rm m}\rangle), \hat \varrho] + \mathcal{L}_\mathrm{f}\hat \varrho +  \mathcal{L}_\mathrm{h}\hat \varrho + \mathcal{L}_\mathrm{c}\hat \varrho\equiv \mathcal{L}(\langle \hat{x}_{\rm m}\rangle)\hat{\varrho}, 
\end{equation}
where 
\begin{equation}
    \hat H(\langle \hat{x}_{\rm m}\rangle) = \sum_{j=1,2,3}\varepsilon_j\hat{p}_j + \Omega_\mathrm{f}\hat a^\dagger \hat a+ f (\hat a^\dagger \hat{\sigma} + \hat a \hat{\sigma}^\dagger) -\sqrt{2}g \hat a^\dagger \hat a \langle \hat{x}_\mathrm{m}\rangle\,.
\end{equation}
This master equation has to be solved together with the equations of motion that govern the mechanical oscillations
\begin{align}
	    \frac{d}{dt} \langle \hat{x}_\mathrm{m}\rangle &= \Omega_\mathrm{m} \langle \hat{p}_\mathrm{m}\rangle - \frac{\gamma_\mathrm{m}}{2}\langle \hat{x}_\mathrm{m}\rangle \,, \label{eq:xm}
    \\
    \frac{d}{dt} \langle \hat{p}_\mathrm{m}\rangle &= -\Omega_\mathrm{m} \langle \hat{x}_\mathrm{m}\rangle - \frac{\gamma_\mathrm{m}}{2}\langle \hat{p}_\mathrm{m}\rangle +\sqrt{2} g\langle \hat a^\dagger \hat a\rangle \,. \label{eq:pm}
\end{align}
In this description, the mechanical oscillator and the cavity only affect each other through their averages, neglecting any correlations between them. We note however that the correlations between the cavity and the emitter are fully taken into account.
With this approximation the system can be numerically solved. Compared to the lasing case 
$f\neq0,\,g=0$ discussed in the previous Section, 
we generically find that the presence of a nonzero optomechanical coupling results in mechanical oscillations that survive in the long-time limit, in the presence of mechanical damping. Moreover, two qualitatively distinct regimes are found that can be characterized by the long-time value of the population inversion $\langle \hat \sigma_z\rangle_\mathrm{ss}=\lim_{T\rightarrow +\infty}T^{-1}\int_{t_0}^{t_0+T} \langle \hat \sigma_z\rangle \mathrm{d}t$. If $\langle \hat \sigma_z\rangle_\mathrm{ss}<0$ the system approaches either a truly stationary state or a state characterized by mechanical oscillations with no appreciable amplitude, while if $\langle \hat \sigma_z\rangle_\mathrm{ss}>0$ the system displays self-sustained mechanical oscillations with large amplitude, which is the regime we are interested in. This confirms that the single-emitter maser acts as the power unit of the clock.

In Fig.~\ref{f:LimitCycle}\,(b) we show the evolution of the expectation values of the two mechanical quadratures in phase space, which confirms that the system settles into a limit cycle. Although a single trajectory is shown, the long-time dynamics converge toward the limit cycle for any choice of initial state in the mechanical phase space. We note that we set $\bar n_\mathrm{c}=\bar n_\mathrm{m}=\bar n_{\mathrm f}=0$ when investigating the clock performance quantitatively. We consider finite temperatures in Sec.~\ref{sec:thermodynamics}.

\subsection{Ideal clock regime}
\label{sec:ideal-clock}

\begin{figure}[t]
\centering
\begin{overpic}[width=0.99\columnwidth]{ 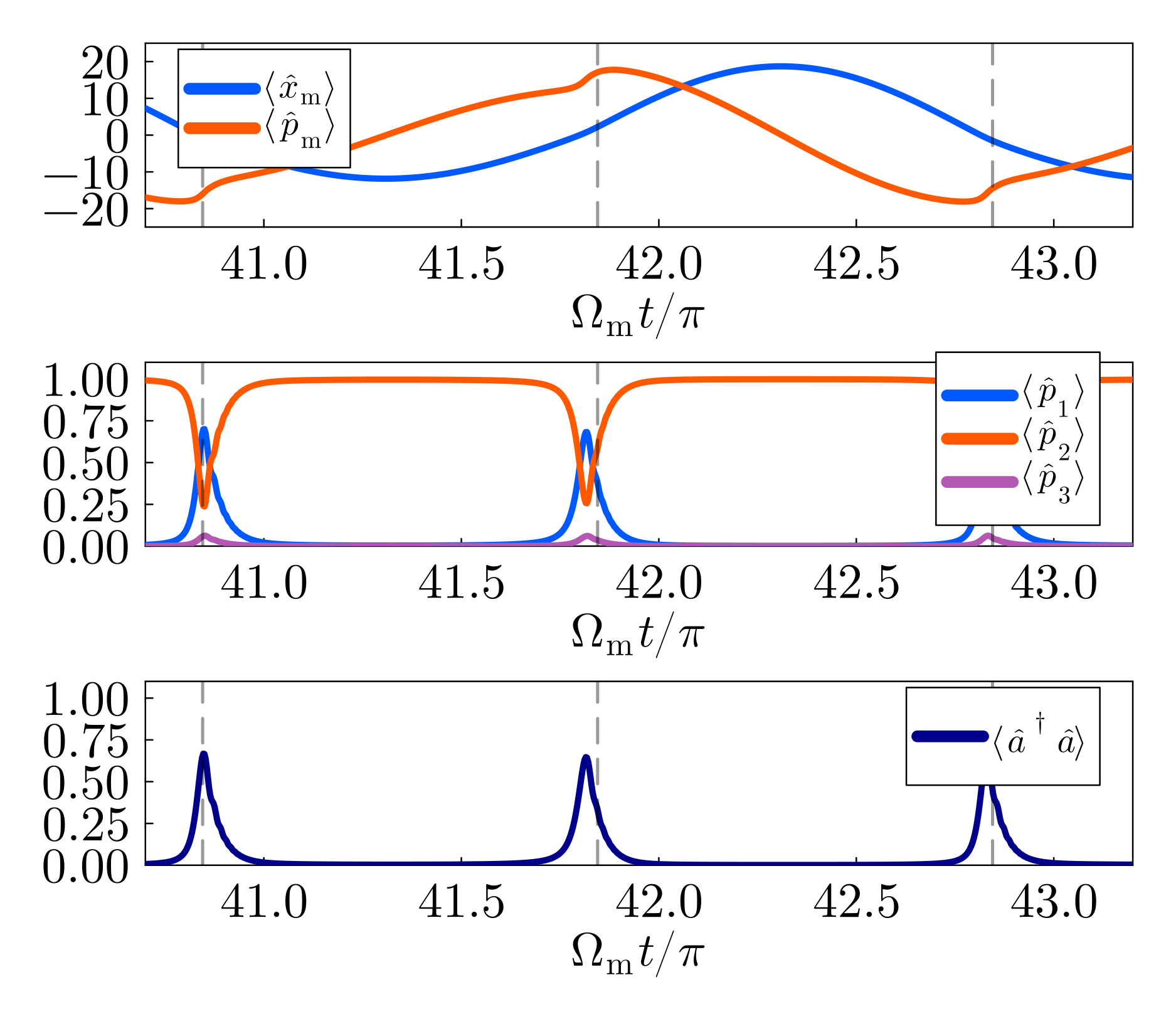}
 \put(30, 78){\large{(a)}}
 \put(20, 50){\large{(b)}}
 \put(15, 24){\large{(c)}}
\end{overpic}

\caption{\label{f:PopCavEvo}  Long time evolution of (a) the average mechanical quadratures, (b) the emitter populations and (c) the mean photon number. Parameters are the same as those used for Fig.~\ref{f:LimitCycle} (b). The dashed gray lines are half a mechanical period apart.}
\end{figure}

From Fig.~\ref{f:LimitCycle}\,(b) we see that the mechanical self-oscillations have a striking appearance, roughly consisting of two semi-circular paths of different amplitudes pieced together. This shape is the result of the escapement action and informs about the clock's operation, as we illustrate below. The evolution of the mechanical quadratures is dynamically correlated with the behaviour of the average emitter populations and photon number, as shown in Fig.~\ref{f:PopCavEvo}\,(a)-(c). The optomechanical pendulum clock goes through a cycle analogous to that of the grandfather clock. Following the four steps of Fig.~\ref{f:LimitCycle}\,(a) we have:

\begin{enumerate}[itemsep=1pt, left=0pt]
\item Starting from maximum negative excursion 
the cavity and lasing transition are far off resonant, the effective detuning being given by $\Delta - \sqrt2 g\langle \hat x_\mathrm{m}\rangle$, cf. Eq.~\eqref{eq:ham} and left panel of Fig.~\ref{f:LimitCycle}\,(a). For simplicity, we take $\Delta=0$ in the following. The mechanical resonator undergoes free motion under the action of the restoring force. The corresponding state of the emitter and the cavity, respectively shown in Fig.~\ref{f:PopCavEvo}\,(b) and (c), shows that the emitter population is fully inverted and the cavity is empty. Full inversion is achieved by ensuring fast relaxation toward the lasing state, i.e., $\gamma_\mathrm{c}\gg \gamma_\mathrm{h}$, and by inhibiting the $\ket{2}\rightarrow\ket{3}$ transition, i.e. $\bar n_\mathrm{c}=0$. 
\item The mechanics sweeps through $\langle\hat x_\mathrm{m}\rangle=0$ and for a brief interval of time the lasing transition becomes resonant with the cavity frequency. The populations $\langle \hat{p}_1\rangle$ and $\langle \hat{p}_2\rangle$ almost fully exchange, and concurrently a photon pulse is released into the cavity. This process takes place over a fraction (roughly 1/10) of a mechanical period. The photon pulse imparts a kick to the mechanical resonator, which experiences a sudden acceleration, see Fig.~\ref{f:PopCavEvo}\,(a). We identify the narrow peak in $\langle \hat{p}_1\rangle$ or $\langle \hat a^\dagger \hat a \rangle$ as the ``tick".
\item Past the resonance both $\langle \hat{p}_1\rangle$ and $\langle \hat a^\dagger \hat a \rangle$ quickly relax to zero, while the mechanical motion  proceeds freely. 
\item The resonance condition is met again and the ``tock" is produced. In this case the photon momentum is opposite to the mechanical momentum, slowing the resonator down, as seen by the sudden reduction in momentum once past the resonance,  see Fig.~\ref{f:PopCavEvo}\,(a). The cycle repeats. 
\end{enumerate}

A distinctive feature of our optomechanical clock is that it proceeds asymmetrically during its limit-cycle motion, reaching larger excursions for positive displacements, to which correspond larger (absolute) values of the momentum, see Fig.~\ref{f:LimitCycle}\,(b). The asymmetry around the equilibrium point translates to more time spent completing the swing in the region of positive displacement than negative ones. This is reflected in non equal spacing between subsequent ticks. 

\begin{figure*}[ht]
 \centering
 \begin{minipage}[b]{0.7\columnwidth}
  \begin{overpic}[width=\linewidth, tics=10]{ phase_space_cond_short.png}
  \put(50, 100){\large{(a)}}
  \end{overpic}
 \end{minipage}
 \hspace{2.5cm}
 \begin{minipage}[b]{0.7\columnwidth}
  \begin{overpic}[width=\linewidth, tics=10]{ 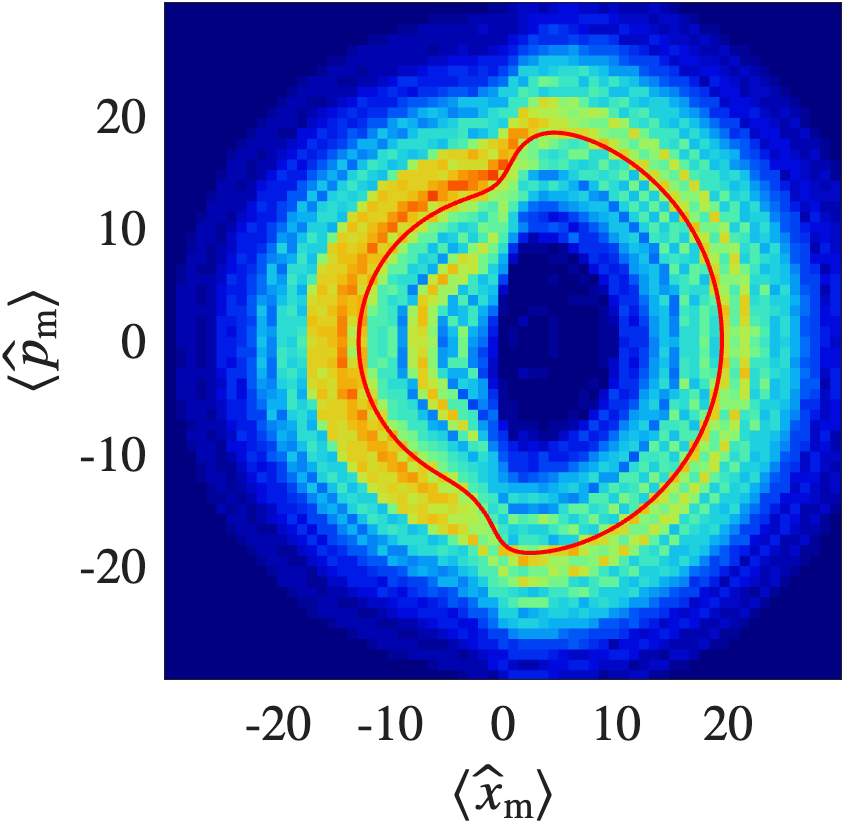}
  \put(50, 100){\large{(b)}}
  \end{overpic}
 \end{minipage}
 \caption{(a) Sample trajectory through phase space of the mechanics for 100 periods, starting at the origin. A similar behavior to the semiclassical analysis is seen, c.f.~Fig.~\ref{f:LimitCycle}. In contrast to the semi-classical analysis, the trajectory is noisy and thus cannot reach a perfect limit cycle. (b) Data from 1000 periods is used to generate the phase-space heatmap, where brighter colors indicate regions of higher occupancy. The red curve denotes the limit cycle associated with the unconditional dynamics, giving perspective to the fluctuations that occur in the conditional case. The other parameters are identical to those in Fig.~\ref{f:LimitCycle}.}
 \label{fig:mechanics_cond}
\end{figure*}

We identify the cycle illustrated above and in Figs.~\ref{f:LimitCycle},~\ref{f:PopCavEvo} with the working regime of an \emph{ideal} clock, which corresponds to an impulsive radiation pressure force, acting over a narrow fraction of the mechanical oscillation, and to the (almost) complete extinction of the mean photon number and emitter ground state population away from resonance; this allows for the unambiguous identification of the ticks and the tocks.
We briefly describe in general terms how this parameter regime is characterized: 
\begin{itemize}[itemsep=1pt, left=0pt]
\item $\Delta=0$, namely the cavity is resonant with the lasing transition; this means that the only way they become off-resonant is dynamical, namely via the optomechanical coupling.  
\item $g \sim f$ and $g,f\gtrsim \kappa$, namely both strong Jaynes-Cummings coupling and optomechanical single-photon strong coupling regime are required. In particular, the second requirement means that it is enough to have a single photon in the cavity to displace the mechanical position (more than its zero point term).
\item $\bar n_\mathrm{c}\sim 0$, namely the emitter cycle becomes uni-directional $\ket{1}\rightarrow \ket{3} \rightarrow \ket{2} \rightarrow \ket{1}$. Once level $\ket{2}$ gets populated, it can only go back to $\ket{1}$ via Rabi oscillations and not incoherent transitions.  
\item $\gamma_\mathrm{c}\gg \gamma_\mathrm{h}$, namely level $\ket{3}$ is barely populated but enables virtual transitions from 1 to 2.  
\item $\gamma_\mathrm{h},\gamma_{\mathrm{c}},\kappa > \Omega_\mathrm{m}/\pi$, ensuring that the cavity is emptied and the population inversion in the emitter is restored before the next tick. In addition, $4g^2/(\kappa+\gamma_{\rm c}\bar{n}_{\rm c}+\gamma_{\rm h}\bar{n}_{\rm h})\gg\Omega_\mathrm{m}$ is required that the photon is emitted into the cavity during the short time the cavity is resonant with the emitter.
\item For the Markovian master equation to be justified, we furthermore require $\omega_{13},\,\omega_{23},\,\omega_{12},\, \Omega_{\rm f} \gg \Omega_{\rm m},\,\kappa,\,\gamma_{\rm c},\,\gamma_{\rm h},\,f,\,g$ and $\Omega_\mathrm{m}\gg\gamma_{\mathrm{m}}$.
\end{itemize}

\subsection{Resolving the ticks}
\label{sec:ticks}
\begin{figure}[b]
\centering
\begin{overpic}[width=\columnwidth]{ 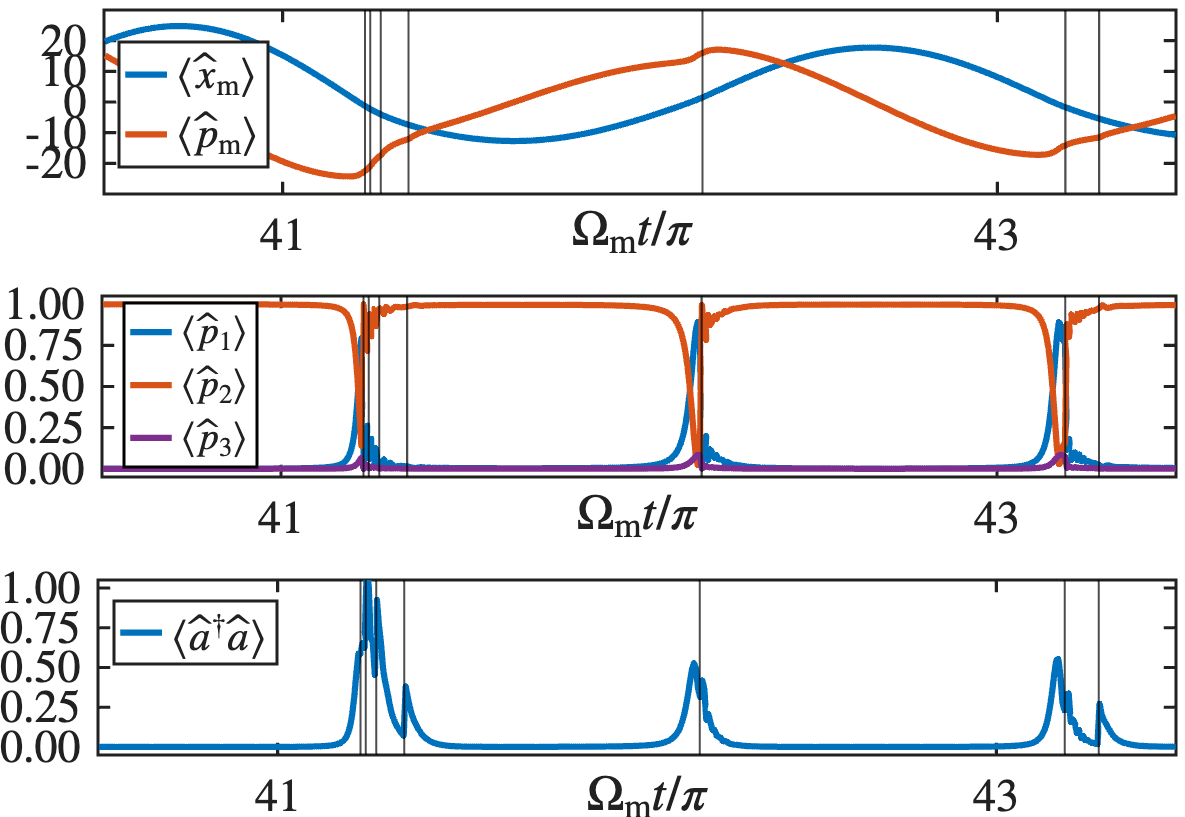}
 \put(9, 68.9){\large{(a)}}
 \put(9, 44.9){\large{(b)}}
 \put(9, 21){\large{(c)}}
\end{overpic}
 \caption{Sample trajectory of (a) quadratures, (b) three-level populations, and (c) cavity occupation. We observe a periodic behavior similar to the classical analysis but noisy, see Fig.~\ref{f:PopCavEvo}. The gray vertical lines denote the times at which ticks occur. During one transition through resonance (one peak in the cavity occupation), multiple ticks may occur. Parameters are the same as in Fig.~\ref{f:LimitCycle}.}
\label{fig:trajectories}
\end{figure}

As illustrated in Fig.~\ref{f:Sketch}\,(b), a convenient way to count the ticks and tocks of the clock is to infer them from the photons emitted into the cold bath. We  assume that the emitted photons can be detected with unit efficiency for simplicity. We also refer to them just as ticks in the following. The ticks are then described by the superoperator
\begin{equation}
\label{eq:jump}
    \mathcal{J}\hat{\varrho} = \gamma_{\mathrm c}(\bar{n}_\mathrm{c}+1)\hat{\sigma}_{23}\hat{\varrho}\hat{\sigma}_{32}.
\end{equation}
To keep track of the ticks, we employ a stochastic master equation~\cite{landi:2024}. To this end, we write the master equation in Eq.~\eqref{eq:MFME} as
\begin{equation}
    \label{eq:masterj}
  \hat \varrho(t+dt)= \hat \varrho(t) + [\mathcal{L}(\langle\hat{x}_{\rm m}\rangle)-\mathcal{J}]\hat\varrho(t)dt+\mathcal{J}\hat\varrho(t)dt.
\end{equation}
This equation can be interpreted as a jump happening during the time increment $dt$ with probability 
\begin{equation}
    p = dt {\mathrm Tr}\{\mathcal{J}\varrho(t)\}.
\end{equation}
To obtain a stochastic master equation, we introduce a stochastic variable $dN$, that takes on the value $dN=1$ with probability $p$ and zero otherwise. We then \textit{unravel} Eq.~\eqref{eq:masterj} as
\begin{equation}
    \label{eq:stochmas1}
    \hat\rho(t+dt) = \frac{(1-dN)}{1-p}\left[\hat\rho(t)+(\mathcal{L}-\mathcal{J})\hat\rho(t) dt\right]+\frac{dN}{p}\mathcal{J}\hat\rho(t) dt,
\end{equation}
where, for ease of notation, we use $\mathcal{L}\equiv\mathcal{L}(\langle\hat{x}_{\rm m}\rangle)$.
Furthermore, we introduced the symbol $\hat{\rho}$ to denote the density matrix conditioned on all previous ticks. In each time-step, a tick may happen $dN=1$, and the density matrix is updated to the second term on the right-hand side of Eq.~\eqref{eq:stochmas1}. If no tick happens, $dN=0$, the density matrix is updated according to the first term on the right-hand side of Eq.~\eqref{eq:stochmas1}.
Averaging Eq.~\eqref{eq:stochmas1} over the different outcomes of $dN$ reproduces Eq.~\eqref{eq:masterj} since $\mathrm{E}[dN\hat{\rho}]=p\hat{\varrho}$. Expanding Eq.~\eqref{eq:stochmas1}, and dropping terms proportional to $dNdt$ (since $p\propto dt$), we arrive at
\begin{equation}
    \label{eq:stochmas2}
    d\hat\rho =\mathcal{L}(\langle\hat{x}_{\rm m}\rangle)\hat\rho dt+(dN-dt\mathrm{Tr}\{\mathcal{J}\hat\rho\})\left(\frac{\mathcal{J}\hat\rho}{\mathrm{Tr}\{\mathcal{J}\hat\rho\}}-\hat\rho\right),
\end{equation}
where $d\hat\rho=\hat\rho(t+dt)-\hat\rho(t)$. This equation explicitly keeps track of the ticks of the clock, since in each time increment $dt$, $dN=1$ denotes a tick, while $dN=0$ denotes no tick. The mechanical quadratures are still described by Eqs.~\eqref{eq:xm} and \eqref{eq:pm} which need to be solved together with Eq.~\eqref{eq:stochmas2}. As illustrated in Fig.~\ref{fig:mechanics_cond} by an example phase-space trajectory for the mechanical oscillator, the stochastic nature of the ticks prevents a perfect limit cycle to establish in contrast to the average evolution illustrated in Fig.~\ref{f:LimitCycle}. Figure \ref{fig:trajectories} illustrates how the emitter and the cavity occupation behave along a trajectory. A tick occurs whenever the emitter emits a photon into the cold bath. As anticipated, the ticks occur when the cavity is in resonance with the emitter. During each of these resonances, the emitter cycles through the states $\ket{1}\rightarrow \ket{3}\rightarrow \ket{2}\rightarrow \ket{1}$, with the transition $\ket{3}\rightarrow \ket{2}$ corresponding to a tick. Importantly, multiple cycles may be completed during the same resonance, resulting in multiple ticks in quick succession. To test the validity of Eq.~\eqref{eq:stochmas2}, we benchmark it against an unraveled version of Eq.~\eqref{eq:FullME} for parameters that result in mechanical oscillations with smaller amplitudes, finding good agreement as shown in App.~\ref{app:benchmark}.

\section{Clock performance}
\label{sec:clockperf}
\begin{figure*}[ht]
    \centering
    \begin{overpic}[width=0.8\textwidth]{ 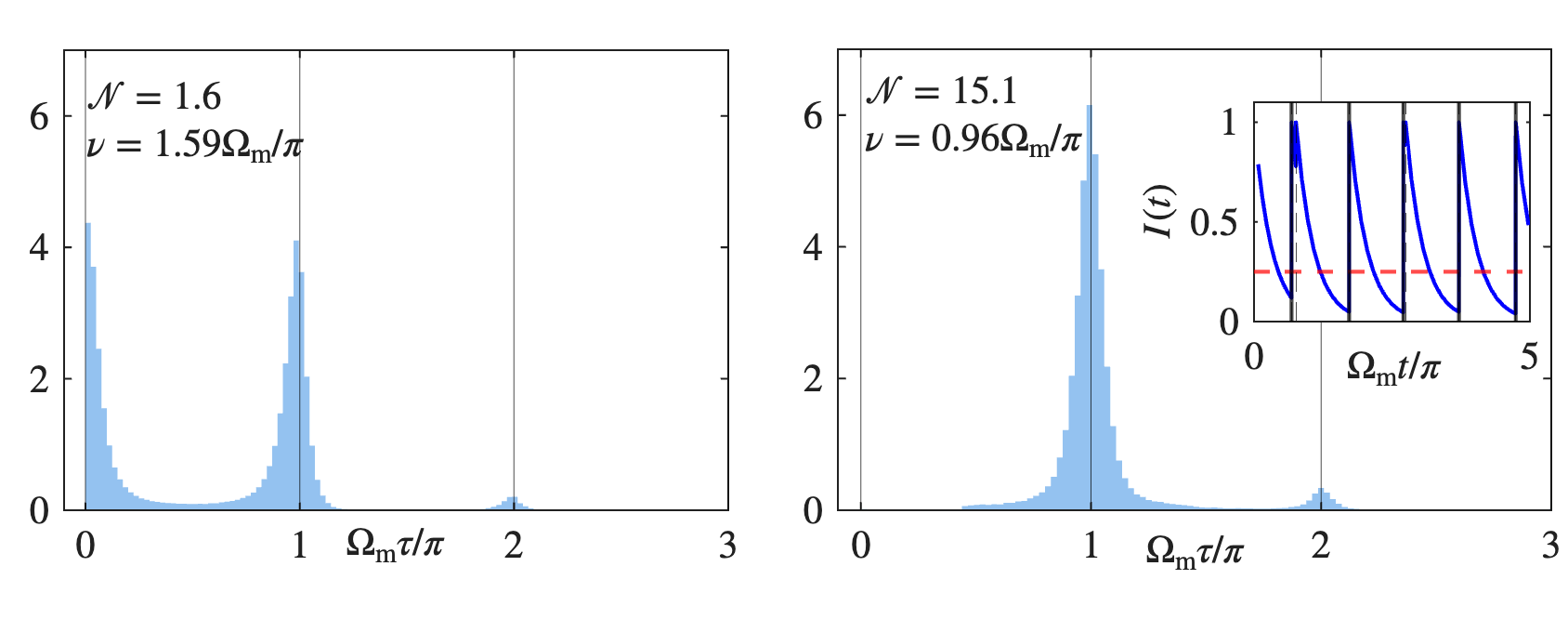}
        \put(5, 38){\large{(a)}}
        \put(55, 38){\large{(b)}}
    \end{overpic}
    \caption{Histogram of ticks (a) before and (b) after filtering. Before filtering, the histogram has a large peak at 
$\tau\simeq 0$ because multiple ticks can occur while the mechanics moves through a single resonance. Filtering effectively removes the first peak in the histogram, considerably improving the accuracy. Inset: Filtering procedure. The dashed vertical lines show the ticks before filtering, the solid vertical lines show the ticks after filtering. The filtering results in a tick whenever the blue line [Eq.~\eqref{eq:current}] crosses the red-dashed line ($I^*$) from below. The parameters are the same as in Fig.~\ref{f:LimitCycle}, with $I^*=0.25$ and the detector bandwidth $\gamma=\Omega_{\rm m}$.}
    \label{fig:histograms}
\end{figure*}
To characterize the clock, we focus in this section on the idealized scenario where all thermal baths, except for the hot one, are at zero temperature. This assumption will be relaxed in Sec.~\ref{sec:thermodynamics}.

\subsection{Waiting times}

We characterize the clock by the waiting times between subsequent ticks. A histogram of such waiting times, the waiting-time distribution, is shown in Fig.~\ref{fig:histograms}\,(a). It consists of three peaks, one close to zero, one close to half a period, and one close to the period of the mechanical oscillator. The peaks close to zero as well as the peak close to a full period are undesired, since a perfect clock has the same waiting time between all subsequent ticks and thus results in a waiting-time distribution with a single, infinitely narrow peak. The peak close to zero is a consequence of the occurrence of multiple ticks during a single resonance, see Fig.~\ref{fig:trajectories}. The small peak at the mechanical period occurs when the emitter does not cycle through the three states during a resonance, making the clock skip a tick. 

To characterize the performance of the clock, we introduce two figures of merit, the accuracy and the resolution. The accuracy is defined as the square of the average waiting time divided by its variance~\cite{erker:2017,meier:2023}
\begin{equation}
    \mathcal{N} = \frac{\langle \tau\rangle^2}{\langle\!\langle \tau^2\rangle\!\rangle}.
\end{equation}
The accuracy may be interpreted as the average number of ticks before the clock is off by one tick (on average). For the histogram Fig.~\ref{fig:histograms}\,(a), the accuracy is very low with $\mathcal{N}\simeq 1.6$. For comparison, a Poisson process results in an exponential waiting-time distribution with unit accuracy~\cite{meier:2023}. The reason for this bad performance are the undesired peaks in the waiting-time distribution.

Another important performance quantifier is the resolution of the clock, given by the inverse of the mean waiting time
\begin{equation}
    \label{eq:resolution}
    \nu = \frac{1}{\langle\tau\rangle},
\end{equation}
which characterizes the shortest time that can be resolved by the clock. Generally, the resolution and the accuracy are not independent and one may increase the accuracy of a clock by reducing its resolution. For instance, counting each $m$-th event in a Poisson process as a tick results in an improvement of the accuracy by a factor of $m$. However, the product $\mathcal{N}\nu$ remains invariant. This trade-off between accuracy and resolution is captured by recently derived upper bounds on the products $\mathcal{N}\nu$~\cite{prech:2024} and $\mathcal{N}\nu^2$~\cite{meier:2023}.

\subsection{Detecting ticks with finite bandwidth}

To overcome the limitation in our clock presented by the peak in the waiting-time distribution at short times, we consider the ticks giving rise to a detector current
\begin{align}\label{eq:current}
    I(t) =  \sum_j e^{-\gamma (t-t_j)} \left[\Theta(t-t_j) - \Theta(t-t_{j+1})\right],
\end{align}
where $t_j$ denotes the time of the $j-th$ original ticks, $\gamma \geq 0$ is a decay rate, and $\Theta$ is the Heaviside theta function. The current in Eq.~\eqref{eq:current} takes the form $e^{-\gamma\tau}$, where $\tau$ is the time since the last tick. It thereby corresponds to a low-pass filter of bandwidth $\gamma$~\cite{wiseman:book}, and thus provides a realistic response of an experimental detection scheme for the ticks. From the detector current, we may then define ticks by the times $I(t)$ crosses a threshold value $I^*$ from below. This is illustrated in the inset of the right panel of Fig.~\ref{fig:histograms}. As shown in Fig.~\ref{fig:histograms}, this results in a histogram that does not exhibit the peak at short waiting times, since $I(t)$ has to relax below $I^*$ before the next tick can be detected. Thus, Eq.~\eqref{eq:current} may also be interpreted as modeling the dead time of a detector.

After this procedure, the accuracy of our quantum pendulum clock increases by an order of magnitude to $\mathcal{N}\simeq 15.1$. 
 We note that this increase in accuracy is not explained by the above-mentioned accuracy vs resolution trade-off, as the resolution merely decreases by less then a factor of two, from $\nu=1.59\,\Omega_{\rm m}/\pi$ to $\nu = 0.96\,\Omega_{\rm m}/\pi$.
The accuracy of our pendulum clock is still much below Ref.~\cite{culhane:2024}, where accuracies of the order of $10^4$ were found for a clock based on an electromechanical architecture. The reason for this is that we explicitly model the detection of the ticks through photo-detection. This results in periods without a tick, resulting in the small peak at $\tau \simeq 2\pi/\Omega_{\rm m}$. In contrast, in Ref.~\cite{culhane:2024}, the ticks are obtained from the continuous trajectory of the mechanical oscillator, which requires a classical description for the pendulum.

The limited accuracy illustrates that our quantum pendulum clock is affected by both thermal and quantum noise. This is very different from a classical, macroscopic pendulum clock, which remains unaffected by thermal fluctuations for a long time.
In Sec.~\ref{sec:classical_limit}, we discuss how such a classical behavior may be approached, by embedding many emitters in a cavity.

\subsection{Allan variance}

As a further means of characterizing the clock, we consider the Allan variance ~\cite{allan1966statistics,vig1999ieee}, which provides a measure of the stability of the clock. To introduce the Allan variance, we define the departure of the reading of our pendulum clock from a perfect clock. Since our quantum pendulum clock produces discrete ticks, it only provides a reading at discrete times. We thus introduce
\begin{equation}
\label{eq:deviation}
X_j  = j\langle \tau\rangle - t_j,
\end{equation}
where $j\langle \tau\rangle$ denotes the reading of the clock, assuming that we assign the average waiting time as the time increment of our clock. Due to the stochastic nature of the clock, its reading will generally deviate from the true time $t_j$ at which the $j$-th tick occurs. We further introduce
\begin{equation}
    Y_j(m) = \frac{X_{j+m}-X_j}{m\langle \tau\rangle},
\end{equation}
which denotes the increase in $X_j$ over $m$ ticks, normalized by the time an ideal clock takes for $m$ ticks. The Allan variance is then defined as
\begin{equation}
\label{eq:allan}
\begin{aligned}
    &\sigma^2_\mathrm{A}(m) = \frac{1}{2}\left\langle [Y_{j+m}(m)-Y_j(m)]^2\right\rangle
    \\& =\frac{1}{2m^2\langle \tau\rangle^2}
\left\langle[t_{j+2m}-2t_{j+m}+t_j]^2\right\rangle.    
\end{aligned}
\end{equation}
Note that after the initial transient dynamics, the Allan variance does not depend on $j$, since it does not matter when we start observation.

\begin{figure}[b]
      \includegraphics[width=\columnwidth]{ 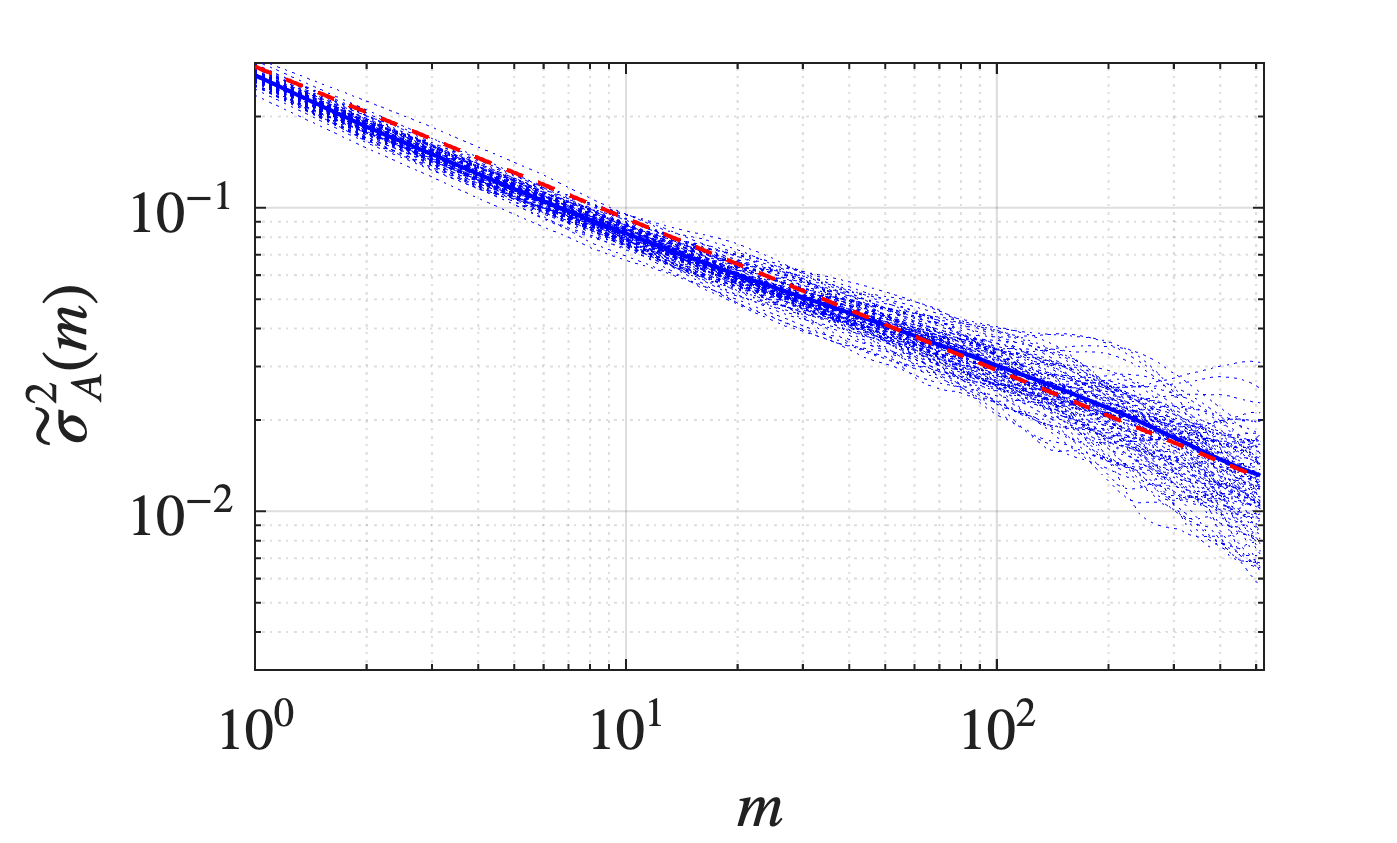}
    \caption{Allan variance after filtering. The light blue curves are obtained from $100$ single trajectories, according to Eq.~\eqref{eq:allantraj}. The thick blue line is their average. The red-dashed line is $1/\mathcal{N}m$, the long-time limit of the Allan variance. Note that for large $m$, the fixed length of the trajectories result in less data points, explaining the larger differences between individual trajectories. Here, we have $L=2028$. The parameters are the same as in Fig.~\ref{fig:histograms}.}
\label{fig:allan}
\end{figure}

\begin{figure*}[!htb]
    \centering
    \begin{overpic}[width=0.95\linewidth]{ 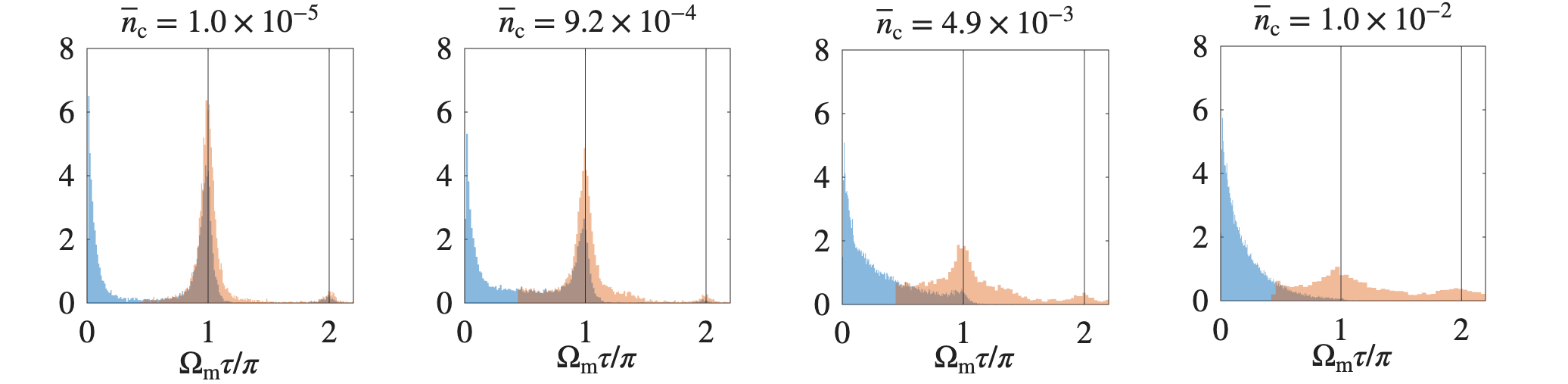}
        \put(13.5, 25.4){\large{(a)}}
        \put(36, 25.4){\large{(b)}}
        \put(60, 25.4){\large{(c)}}
        \put(85, 25.4){\large{(d)}}
    \end{overpic}
    \caption{Waiting time distributions for unfiltered (blue) and filtered (brown) ticks. From (a) to (d), the cold temperature increases. As the temperature rises, spurious ticks contribute to a background in the distributions, eventually spoiling the periodic behavior of the clock. Aside from $\bar n_{\rm c}$ and $T_{\rm m}=T_{\rm f}=T_{\rm c}$, the rest of the parameters are the same as in Fig.~\ref{fig:histograms}.}
    \label{fig:td_entropy}  
\end{figure*}

For large $m$, the Allan variance simplifies since we can introduce two approximately identical and independently distributed random variables $\bar{\tau}_j$ and $\bar{\tau}_{j+m}$ with
\begin{equation}
    \bar{\tau}_j = (t_{j+m}-t_j)/m,
\end{equation}
which is the average of $m$ subsequent waiting times. For sufficiently large $m$, almost all waiting times in $\bar{\tau}_j$ are uncorrelated from almost all waiting times in $\bar{\tau}_{j+m}$.
The Allan variance can then be written as
\begin{equation}
\label{eq:allan_long}
    \sigma_\mathrm{A}^2(m) = \frac{\left\langle( \bar{\tau}_{j+m}- \bar{\tau}_j)^2\right\rangle}{2\langle \tau\rangle^2}
=\frac{\langle\!\langle \bar{\tau}^2_j\rangle\!\rangle}{\langle \tau\rangle^2}=\frac{1}{m\mathcal{N}},
\end{equation}
where we used that $\langle \bar{\tau}_j\rangle=\langle \tau\rangle$ and for sufficiently large $m$, $\langle\!\langle \bar{\tau}^2_j\rangle\!\rangle=\langle\!\langle \tau^2\rangle\!\rangle/m$ due to the central limit theorem. The Allan variance, together with its long-time limit, is illustrated in Fig.~\ref{fig:allan}. The fact that the Allan variance agrees with its long-time limit already at short times implies that correlations between consecutive ticks are small. This is indeed the case, as shown in App.~\ref{app:correlations}.

Evaluation of the Allan variance~\eqref{eq:allan} requires an ensemble average. Here, we use an estimate for it based on a single trajectory of $L$ ticks, the so-called overlapping Allan variance~\cite{vig1999ieee}
\begin{align}
\label{eq:allantraj}
    {\tilde\sigma}_{\rm A}^2(m)  = \frac{1}{2m^2\langle\tau\rangle^2 (L-2m)}\sum_{j=1}^{L-2m}(t_{j+2m} - 2 t_{j+m} + t_j)^2,
\end{align}
which is similar to Eq.~\eqref{eq:allan}, except that the ensemble average is replaced by averaging over the time series. 

\section{Thermodynamic analysis}
\label{sec:thermodynamics}
\begin{figure*}[ht]
    \centering
    \begin{overpic}[width=.7\linewidth]{ 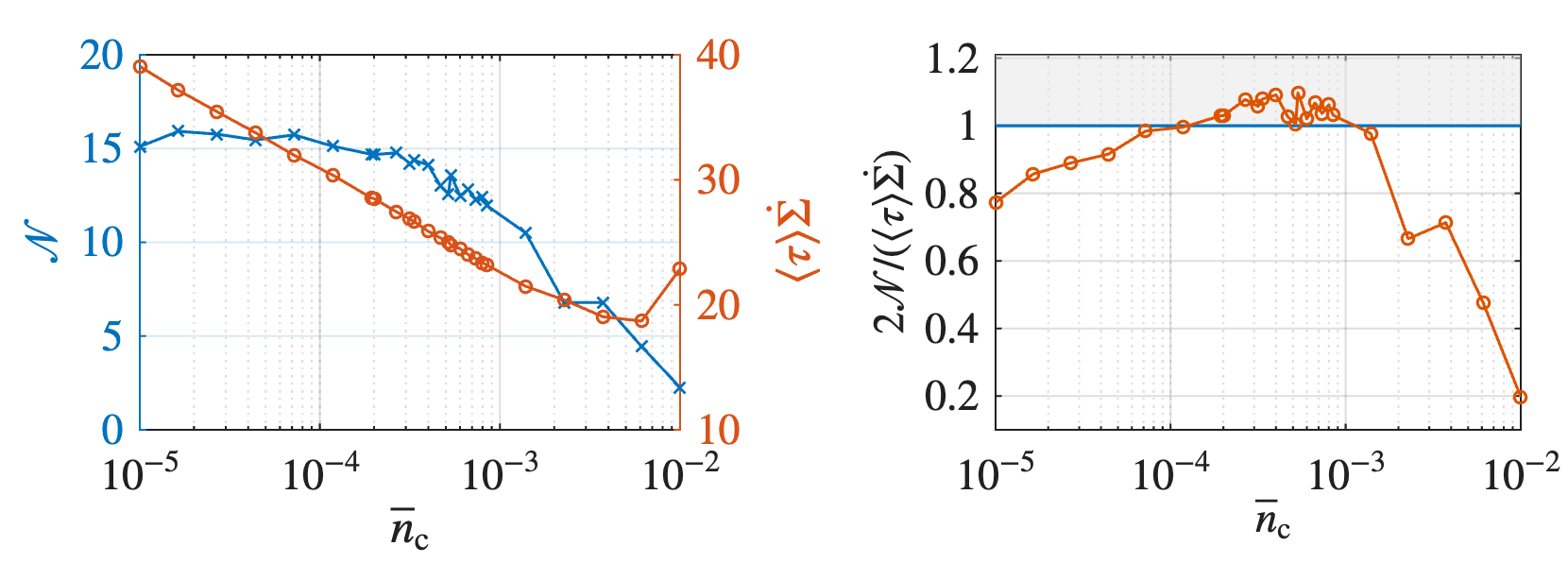}
        \put(10, 36){\large{(a)}}
        \put(64, 36){\large{(b)}}
    \end{overpic}
    \caption{Thermodynamic uncertainty relation (TUR). (a) Filtered accuracy (blue crosses), and entropy production per tick (red circles) as a function of the cold temperature. As expected, a better clock dissipates more strongly. The increase of the entropy production per tick at high $\bar{n}_{\rm c}$ is due to the increase of the average waiting time $\langle\tau\rangle$, see also Fig.~\ref{fig:td_entropy}. (b) The TUR ratio (using the filtered data), where the shaded area represents violations of the classical TUR. At sufficiently low temperatures the data shows a violation of the bound. Parameters are the same as in Fig.~\ref{fig:td_entropy}.}
    \label{fig:tur}
\end{figure*}
In this section we examine the trade-off between precision and dissipation for the pendulum clock. In our setup, the first law of thermodynamics (in steady state) reads
\begin{equation}
    \label{eq:firstlaw}
     J_{\rm f}+ J_{\rm m} +J_{\rm c} + J_{\rm h} = 0,
\end{equation}
where $J_\alpha$ denotes the heat that enters bath $\alpha$. The second law of thermodynamics is given by
\begin{align}\label{eq:2ndlaw}
    \dot\Sigma = \frac{J_{\rm f}}{T_{\rm f}}+ \frac{J_{\rm m}}{T_{\rm m}}+\frac{J_{\rm c}}{T_{\rm c}}+\frac{J_{\rm h}}{T_{\rm h}} \geq 0,
\end{align}
where $\dot\Sigma$ denotes the entropy production.
In the following, we assume that all temperatures except for the hot one are equal, i.e., $T_{\rm m}=T_{\rm f}=T_{\rm c}$. 
In this case, we can use the first law to eliminate all heat flows except the one into the hot bath from the second law, resulting in
\begin{align}
    \dot\Sigma = J_{\rm h}\left( \frac{1}{T_{\rm h}} - \frac{1}{T_{\rm c}} \right) \geq 0.
\end{align}

Since master equations provide an approximation to the underlying microscopic dynamics of system and environment, it is not always straightforward to compute heat currents in a way that respects the laws of thermodynamics, see for instance Refs.~\cite{levy:2014,hofer:2017,potts:2021}. For our model, we can exploit the separation of energy scales mentioned in the last bullet point in Sec.~\ref{sec:ideal-clock}, see also Eq.~\ref{eq:sepens}, to obtain a good approximation for the entropy production. Since $\omega_{13}\gg f$, the heat carried by photons exchanged with the hot bath is approximately equal to $\omega_{13}$ and we may write
\begin{align}
\label{eq:heathot}
    J_{\rm h} = -{\rm Tr}\{\hat{H}_0{\cal L}_{\rm h} \hat\varrho_{\rm tot}\} = \gamma_{\rm h}\omega_{13}\left[ (\bar n_{\rm h}+1)\average{\hat p_3} - \bar n_{\rm h} \average{\hat p_1} \right].
\end{align}
For a more detailed justification of the first and second laws of thermodynamics in our master equation, see App.~\ref{app:thermodynamics}. 

To obtain a finite entropy production, we allow for a finite cold temperature $T_c$. As we increase the cold temperature (keeping $T_{\rm m}=T_{\rm f}=T_{\rm c}$), the entropy production reduces. However, the clock also deteriorates because now it is possible to absorb a photon from the cold bath, resulting in a transition from state $|2\rangle$ to state $|3\rangle$. Unfiltered ticks are still defined by the photons emitted to the cold bath. This implies that the clock may now tick even out of resonance, purely by exchanging photons with the cold bath resulting in transitions $|2\rangle\rightarrow|3\rangle\rightarrow |2\rangle$. This is illustrated in  Fig.~\ref{fig:td_entropy}. As the cold temperature increases, the desired peak in the waiting-time distribution at half the mechanical period broadens and shrinks.

We now turn to the thermodynamic uncertainty relation, which for clocks takes on the form~\cite{meier:2024,culhane:2024}
\begin{equation}
    \label{eq:tur}
    \mathcal{N}\leq \frac{1}{2}\langle\tau\rangle\dot{\Sigma}.
\end{equation}
While this inequality bounds the accuracy of classical Markovian clocks, it can be overcome in our pendulum clock as illustrated in Fig.~\ref{fig:tur}. As the cold temperature is decreased, we find that both the entropy production as well as the accuracy increase. The accuracy increases faster, resulting in a violation of the TUR, before it plateaus. Once the cold temperature becomes sufficiently small to suppress unwanted transitions, a further decrease only increases dissipation but does not benefit the accuracy. Our clock does therefore not allow to keep increasing the accuracy by increasing the dissipation. This is a consequence of the small dimensionality of the emitter that acts as the escapement of our clock.

We note that only the filtered ticks result in a violation of the TUR. Here, we consider the filtering as a (time-independent) post-processing of the accessible current $I(t)$, similar in spirit to only counting every $m$-th tick in a Poissonian clock. Investigating the entropy production required for such data-processing is an interesting avenue left for future work.

\section{Multiple emitters and the classical limit}
\label{sec:classical_limit}
\begin{figure*}
    \centering
    \includegraphics[width=.7\linewidth]{ 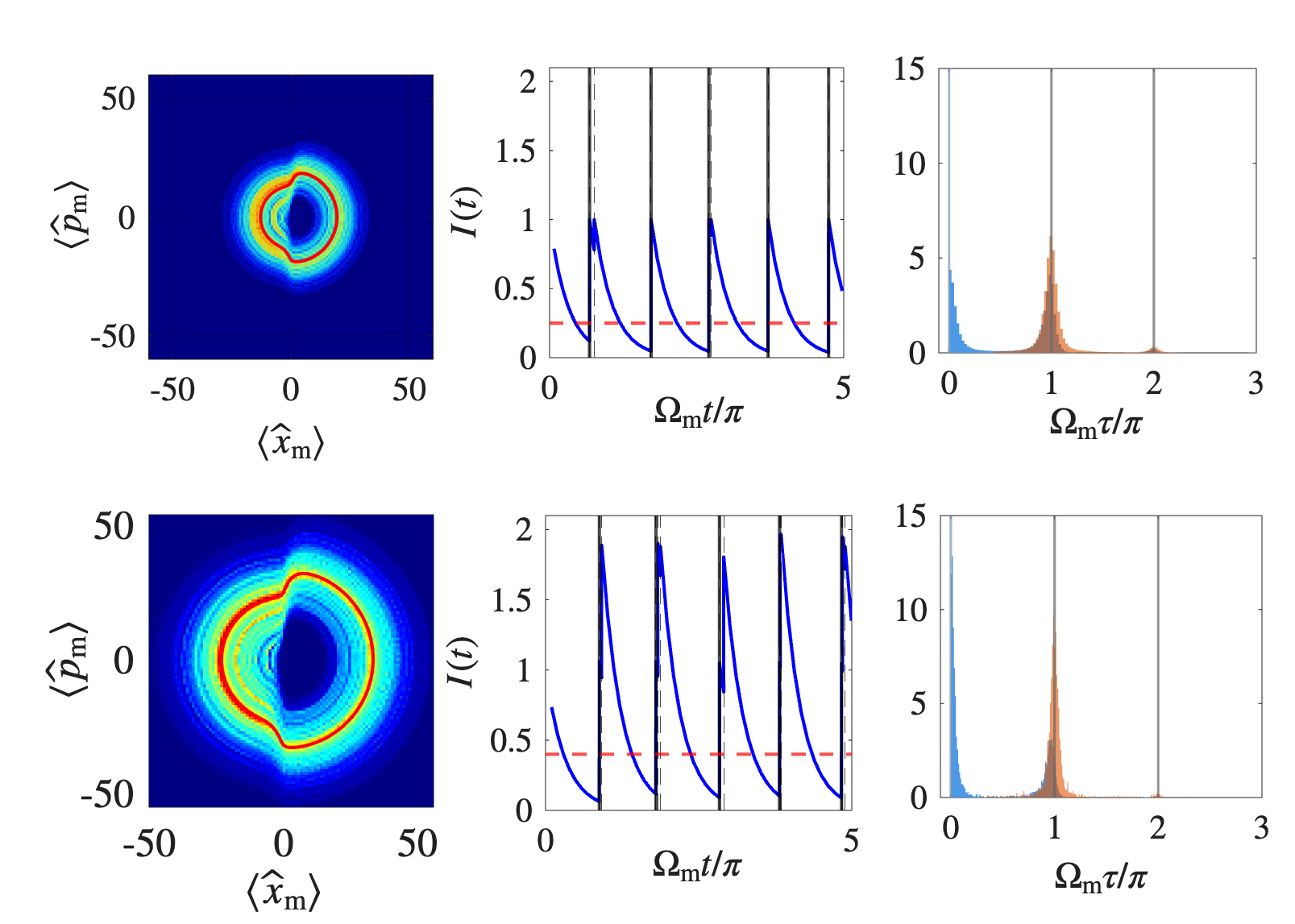}
    \caption{Top: $M=1$ with the following performance metrics. Unfiltered: $\nu = 1.59\Omega_{\rm m}/\pi$, ${\cal N }= 1.6 $. Filtered: $\nu = 0.96\Omega_{\rm m}/\pi$, ${\cal N} = 15.1$. Entropy production: $\dot{\Sigma}\langle\tau\rangle = 39.1$.  Bottom: $M=2$ with the following performance metrics. Unfiltered: $\nu = 2.58\Omega_{\rm m}/\pi$, ${\cal N }= 0.7 $. Filtered: $\nu = 0.99\Omega_{\rm m}/\pi$, ${\cal N} = 35.6$. Entropy production: $\dot{\Sigma}\langle\tau\rangle = 58.6$. Here, we set $\bar n_{\rm c} = 10^{-5}$, $T_{\rm m}=T_{\rm f}= T_{\rm c}$, $\gamma=\Omega_{\rm m}$, and the filter threshold $I^*=M/(M+3)$. The rest of the parameters are the same as in Fig.~\ref{f:LimitCycle}.}
    \label{fig:M1_vs_M2}
\end{figure*}
In order to achieve a limit where our quantum mechanical pendulum clock becomes noiseless and effectively behaves like a classical clock, we consider multiple emitters instead of just one being embedded in the cavity [see Fig. \ref{f:Sketch} c]. By increasing the number of emitters from one to many, we can investigate the quantum-to-classical transition. To this end, we consider the stochastic master equation in Eq.~\eqref{eq:stochmas2} with the modified Hamiltonian
\begin{equation}
\label{eq:hamM}
\begin{aligned}
    \hat H(\langle\hat{x}_{\rm m}\rangle)=&\sum_{j=1}^M\sum_{k=1}^3\varepsilon_k\hat{p}_k^j  + f\sum_{j=1}^M(\hat a^{\dagger} \hat\sigma_{12}^j + \hat a\hat\sigma_{21}^j)\\&+ \Omega_{\rm f} \hat a^{\dagger} \hat a - \sqrt{2}g\hat a^{\dagger}\hat a\langle\hat x_\mathrm{m}\rangle,
\end{aligned}    
\end{equation}
the dissipators
\begin{equation}
\begin{aligned}
&\mathcal{L}_\mathrm{h}=\gamma_\mathrm{h} (\bar n_\mathrm{h}+1)\sum_{j=1}^M\mathcal{D}[\hat{\sigma}^j_{13}] +\gamma_\mathrm{h} \bar n_\mathrm{h}\sum_{j=1}^M\mathcal{D}[\hat{\sigma}_{31}^j]  \,,\\&
\mathcal{L}_\mathrm{c}=\gamma_\mathrm{c} (\bar n_\mathrm{c}+1)\sum_{j=1}^M\mathcal{D}[\hat{\sigma}^j_{23}] +\gamma_\mathrm{c} \bar n_\mathrm{c}\sum_{j=1}^M\mathcal{D}[\hat{\sigma}_{32}^j] \,,
\end{aligned}
\end{equation}
and the jump operator
\begin{equation}
\label{eq:jumpM}
    \mathcal{J}\hat{\varrho} = \gamma_{\mathrm c}(\bar{n}_\mathrm{c}+1)\sum_{j=1}^M\hat{\sigma}^j_{23}\hat{\varrho}\hat{\sigma}^j_{32}.
\end{equation}
Here $j$ labels the emitters and there is a total of $M$ emitters in the cavity. The projector $\hat{p}_k^j$ projects on the $k$-th energy eigenstate of the $j$-th emitter, and $\hat\sigma_{kl}^j$ denotes the transition operator for $l\rightarrow k$ of emitter $j$. For simplicity, we consider all the emitters to be identical and coupled to the cavity and to the thermal baths with equal coupling strength. According to Eq.~\eqref{eq:jumpM}, we consider each photon emitted to the cold bath as a tick.
Since all the emitters are identical, we may characterize them by the operators
\begin{equation}
    \label{eq:atomops}
    \hat q_k = \frac{1}{M}\sum_{j=1}^M\hat{p}_k^j,\hspace{1cm}\hat\sigma = \frac{1}{M}\sum_{j=1}^M\hat\sigma_{12}^j,
\end{equation}
and we note that $\langle \hat{p}_k^j\rangle=\langle \hat q_k\rangle$ as well as $\langle \hat \sigma_{12}^j\rangle = \langle \hat \sigma\rangle$ for all $j$.

Due to the multiple emitters in the cavity, more photons will be emitted per resonance. Indeed, the probability for a tick to happen scales linearly with $M$
\begin{equation}
\label{eq:probM}
    p = dt {\mathrm Tr}\{\mathcal{J}\varrho(t)\} = dt\gamma_{\mathrm c}(\bar{n}_\mathrm{c}+1) M \langle\hat q_3\rangle.
\end{equation} This results in larger limit cycles, which makes to clock more stable as shown below. Since many photons can be emitted into the cold bath during each mechanical period, filtering is even more important for $M>1$. To this end, we consider the current
\begin{align}\label{eq:currentM}
    I(t) =  \sum_j e^{-\gamma (t-t_j)} \left[\Theta(t-t_j) - \Theta(t-t_{j+M})\right],
\end{align}
which recovers Eq.~\eqref{eq:current} for $M=1$. As above, a tick is defined when $I(t)$ crosses a threshold value $I^*$ from below. To compute the entropy production, Eq.~\eqref{eq:2ndlaw} may still be employed. However, the heat current now reads
\begin{equation}\label{eq:heathotM}
    J_{\rm h}  =-{\rm tr}\{\hat{H}_0\mathcal{L}_{\rm h} \hat\varrho_{\rm tot}\}
     = M\gamma_{\rm h}\omega_{13}\left[ (\bar n_{\rm h}+1)\average{\hat q_3} - \bar n_{\rm h} \average{\hat q_1} \right].
\end{equation}

\subsection{Two emitters}
We start by considering two emitters. As illustrated in Fig.~\ref{fig:M1_vs_M2}, the limit cycle of the mechanics indeed becomes larger. Furthermore, as there are typically two photons emitted per resonance, the filtered current in Eq.~\eqref{eq:currentM} develops larger peaks compared to $M=1$, which results in a narrower distribution of the filtered ticks. At the same time, events where no tick is recorded during a resonance are suppressed compared to $M=1$. As a result, the accuracy increases by about a factor about two from $\mathcal{N}\approx15.1$ ($M=1$) to $\mathcal{N}\approx35.6$ ($M=2$). The entropy production per tick increases from $\dot{\Sigma}\langle\tau\rangle \approx 39.1 $ ($M=1$) to $\dot{\Sigma}\langle\tau\rangle \approx$ 58.6 ($M=2$). Here we consider a cold temperature such that $\bar n_{\rm c} = 10^{-5}$. In this case, a single emitter does not result in TUR violations while two emitters do, with $2\mathcal{N}/(\dot{\Sigma}\langle\tau\rangle)\approx 1.2 $.

\begin{figure*}[t]
    \centering
   
    \begin{overpic}[width=0.95\linewidth]{ 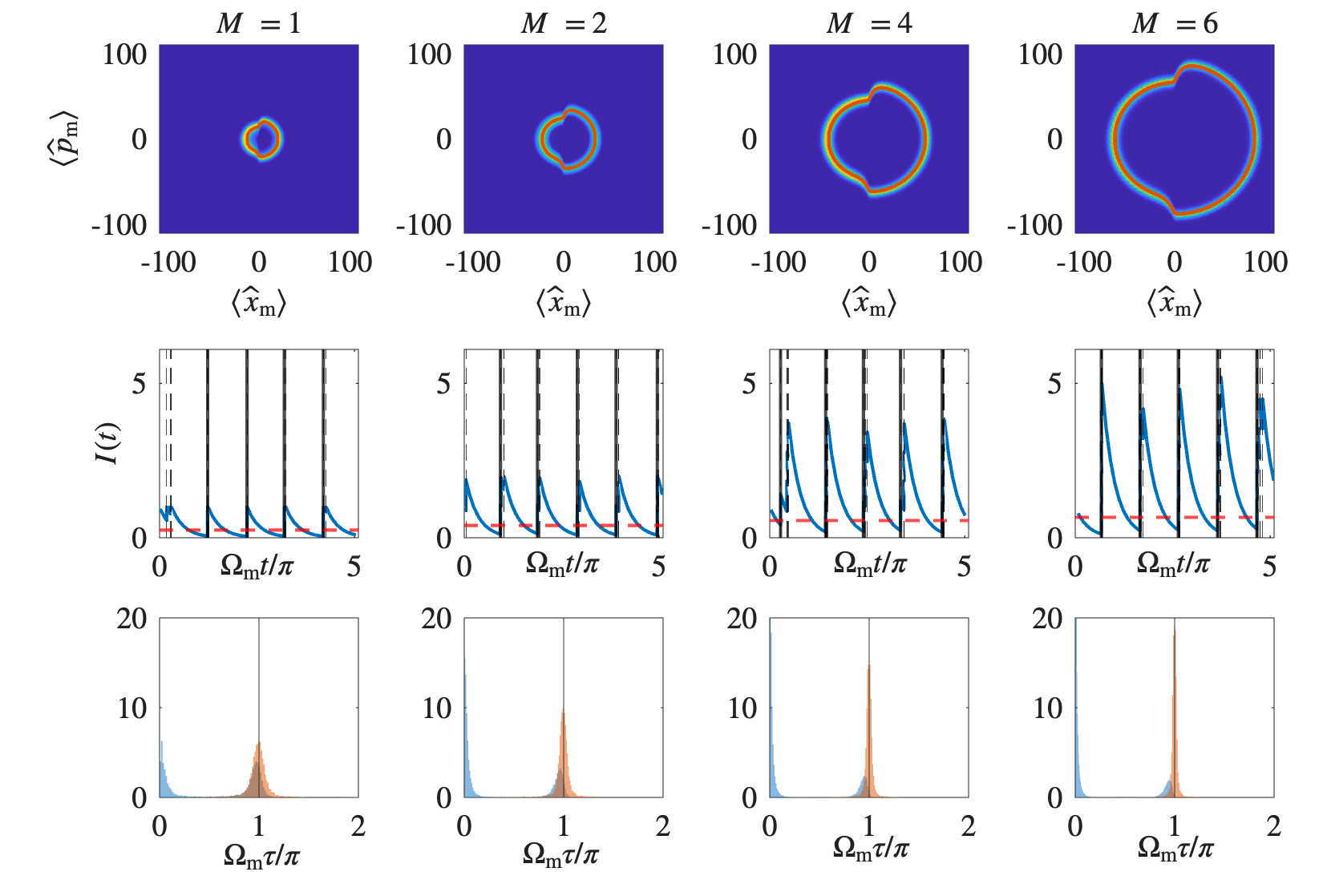}
        \put(1, 64){\large{(a)}}
        \put(1, 39){\large{(b)}}
        \put(1, 20){\large{(c)}}
    \end{overpic}
    \caption{Quantum pendulum clock with multiple emitters $M\in\{1, 2, 4, 6\}$, from left to right, respectively. As $M$ increases, the pendulum clock approaches an ideal behavior. (a) Phase-space densities of the mechanical quadratures (obtained as in Fig.~\ref{fig:mechanics_cond}). As $M$ increases, relative fluctuations around the mean decrease. (b) Current with unfiltered (filtered) ticks illustrated by dashed (solid) lines. The filter threshold is set to $I^*=M/(M+3)$.
    (c) Unfiltered (blue) and filtered (brown) tick histograms. As $M$ increases, the filtered waiting time is narrowly distributed around the average. Parameters are the same as in Fig.~\ref{fig:M1_vs_M2}. The discrepancy between the plots for $M=1,2$ and Fig.~\ref{fig:M1_vs_M2} is due to the additional mean-field approximation.}
    \label{fig:Mclock}
\end{figure*}

\subsection{Many emitters}{\label{sec:M2}}
As $M$ increases further, the Hilbert space dimensions required to simulate the clock become prohibitively large. For this reason, we employ an additional mean-field approximation on the emitters. To this end, we derive equations of motion for the relevant expectation values from the master Eq.~\eqref{eq:stochmas2} with the modified Hamiltonian, dissipators, and jump operator given in Eqs.~\eqref{eq:hamM}-\eqref{eq:jumpM}. We may then obtain a set of closed, nonlinear, stochastic differential equations by employing the factorization
\begin{equation}
\begin{aligned}
&\langle \hat{a}^\dagger\hat{a}\hat{p}_k^j\rangle \rightarrow \langle \hat a^\dagger \hat a\rangle \langle \hat p_k^j\rangle,\hspace{1cm}
    \langle \hat{p}_l^i \hat p_{k}^j\rangle \rightarrow \langle \hat{p}_l^i\rangle\langle \hat p_{k}^j\rangle,\\&
    \langle\hat{p}_l^i \hat \sigma_{rs}^j\rangle \rightarrow \langle \hat{p}_l^j\rangle\langle \hat \sigma_{rs}^j\rangle,\hspace{1cm}
    \langle\hat{\sigma}_{lk}^i \hat \sigma_{rs}^j\rangle \rightarrow \langle \hat{\sigma}_{lk}^i \rangle\langle \hat \sigma_{rs}^j\rangle,
    \end{aligned}
\end{equation}
for $i\neq j$. In this approximation, only the average values, the mean field, of each emitter affects the cavity. We expect this approximation to become more precise as $M$ increases. A similar approach, based on factorizing conditional averages, was recently introduced in Ref.~\cite{yuan:2022} and employed in~\cite{Huihui2024, Nadolny2025} for continuous measurements of optical quadratures. We find the set of coupled equations
\begin{widetext}
\begin{subequations}
\begin{align}
\label{eq:eqsMi}
    d\left\langle \hat{a}^\dagger\hat{a} \right\rangle  &= 2fMdt {\rm Im} \left\langle \hat{a}^\dagger \hat{\sigma} \right\rangle + \kappa dt\left( \bar n_{\rm f} - \left\langle \hat{a}^\dagger\hat{a} \right\rangle \right) ,\\
\label{eq:asM}
    d\left\langle \hat{a}^\dagger \hat{\sigma} \right\rangle  &=  dt\left[i\Delta  -\frac{1}{2}(\kappa+\gamma_{\rm h}\bar n_{\rm h} + \gamma_{\rm c} \bar n_{\rm c})\right]\left\langle \hat{a}^\dagger \hat{\sigma} \right\rangle -i\sqrt{2}g dt\left\langle \hat{a}^\dagger \hat{\sigma}\right\rangle \langle\hat{x}_\mathrm{m}\rangle  + ifdt\left[\langle \hat{q}_2\rangle+ \left\langle \hat{a}^\dagger\hat{a}\right\rangle (\langle\hat{q}_2\rangle - \langle\hat{q}_1\rangle) \right]\\ 
& - \left[\frac{dN}{M} - dt \langle \hat{q}_3\rangle \gamma_{\rm c}(\bar n_{\rm c} + 1)\right]\left\langle \hat{a}^\dagger \hat{\sigma} \right\rangle,\\
\label{eq:q1}
     d\langle \hat{q}_1\rangle  & =2fdt {\rm Im} \left\langle \hat{a}^\dagger \hat{\sigma} \right\rangle + \gamma_{\rm h} dt(\bar n_{\rm h} +1)\langle \hat{q}_3\rangle - \gamma_{\rm h} dt\bar n_{\rm h} \langle \hat{q}_1\rangle 
    - \langle \hat{q}_1\rangle\left[\frac{dN}{M} - dt \langle \hat{q}_3\rangle \gamma_{\rm c}(\bar n_{\rm c} + 1)\right], \\
\label{eq:q2}
    d\langle \hat{q}_2\rangle  &= -2f dt{\rm Im} \left\langle \hat{a}^\dagger \hat{\sigma} \right\rangle + \gamma_{\rm c}dt(\bar n_{\rm c} +1)\langle \hat{q}_3\rangle - \gamma_{\rm c}dt\bar n_{\rm c} \langle \hat{q}_2\rangle 
    +  (1-\langle \hat{q}_2\rangle)\left[\frac{dN}{M} - dt \langle \hat{q}_3\rangle \gamma_{\rm c}(\bar n_{\rm c} + 1)\right].
\end{align}
\end{subequations}
\end{widetext}
\begin{figure*}[tbp!]
    \centering
    \begin{overpic}[width=.85\columnwidth]{ 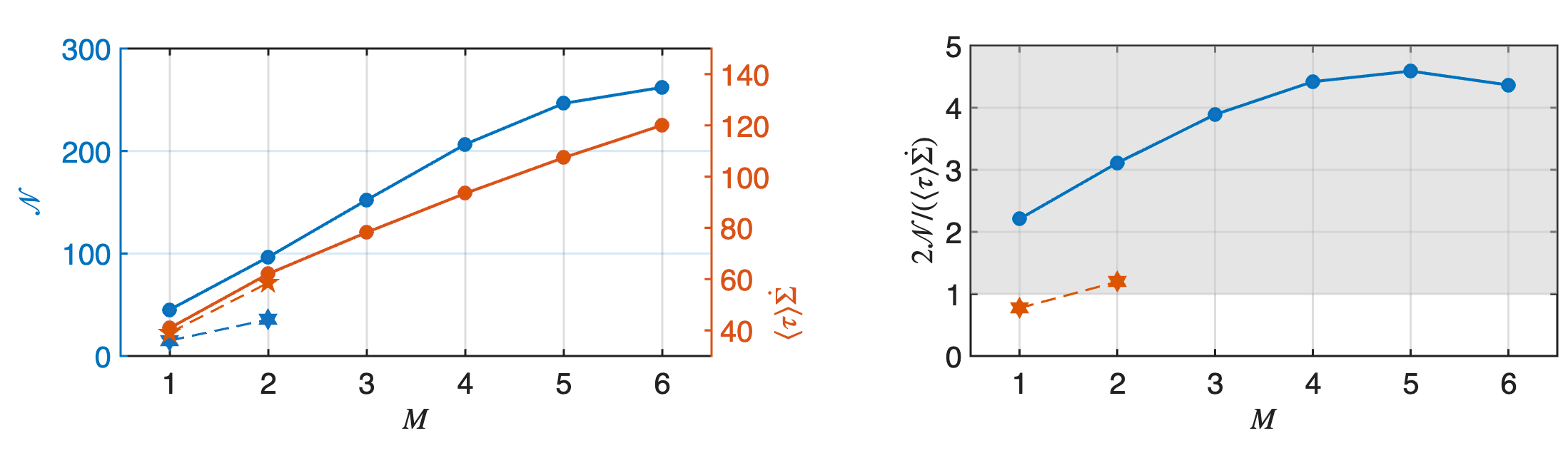}
        \put(8., 30){\large(a)}
        \put(62.2, 30){\large(b)}
    \end{overpic}
    \caption{Accuracy vs. dissipation trade-off for the quantum pendulum clock with multiple emitters. (a) The clock's accuracy and entropy production rate for different values of $M$. (b) The corresponding TUR violations. The clock maintains TUR violations for multiple emitters. As discussed in the main text, these TUR violations are expected to take on a constant value in the limit of infinite emitters, $M\rightarrow \infty$. Dots are obtained from Eqs.~\eqref{eq:eqsMi}-\eqref{eq:q2}, stars are obtained without the additional mean-field approximation. Parameters are the same as in Fig.~\ref{fig:M1_vs_M2}.}
    \label{fig:tdMclock}
\end{figure*}
We note that these averages denote conditional averages, conditioned on all previous ticks of the clock. The mechanical quadratures are still described by Eqs.~\eqref{eq:xm} and \eqref{eq:pm} which need to be solved together with Eqs.~\eqref{eq:eqsMi}-\eqref{eq:q2}. Compared to a single emitter, $M=1$, multiple emitters change the dynamics in two ways. First, the number of photons emitted into the cavity increases by $M$. This can be seen in the first term of the right-hand side of Eq.~\eqref{eq:eqsMi}. Second, the stochastic terms are now proportional to $dN/M$. Whenever a jump happens, this term is thus suppressed by $M$. At the same time, more jumps happen as mentioned above.
As we show in App.~\ref{app:Mnoise}, these two effects combined result in a scaling of the noise terms as $1/\sqrt{M}$, such that we indeed find a noiseless clock for large $M$. Importantly, even the observation of the ticks is possible without noise in this limit.

The time-evolution of our quantum pendulum clock with multiple emitters is illustrated in Fig.~\ref{fig:Mclock}. As $M$ increases, the amplitude of the mechanical oscillator increases and the relative fluctuations around the limit cycle (obtained by dropping the noise terms) decrease.
The histograms of the filtered ticks shown in Fig.~\ref{fig:Mclock} further confirm that the clock approaches an ideal behavior as the waiting-time distribution approaches a single, infinitely narrow peak at the mean waiting time. From the current $I(t)$, we can see that the clock tends towards emitting short bursts of multiple photons twice during the mechanical period. As discussed in App.~\ref{app:Mnoise}, these bursts become deterministic in the large $M$ limit and the pendulum clock is no longer affected by thermal (and quantum) noise.

As expected, the increased accuracy that is obtained by increasing $M$ comes at the cost of an increased entropy production, see Fig.~\ref{fig:tdMclock}. 
From Eq.~\eqref{eq:heathotM}, we anticipate a linear scaling $\dot{\Sigma}\propto M$. As discussed in App.~\ref{app:Mnoise}, the accuracy is also anticipated to scale linearly in the number of emitters $\mathcal{N}\propto M$. We thus expect the TUR violations to persist in the limit of large $M$, where the clock behaves like a macroscopic, classical pendulum clock. For a number of emitters up to $M=6$ this is indeed what we observe, see Fig.~\ref{fig:tdMclock}. For larger values of $M$, the simulation of Eqs.~\eqref{eq:eqsMi}-\eqref{eq:q2} becomes too costly and other theoretical approaches are required. Our findings are in contrast to Ref.~\cite{Gopal_2024_electronicclock}, where TUR violations were only found in the nanoscopic, not in the macroscopic regime.

We note that while the additional mean-field approximation has a small effect on the entropy production, results in considerably higher accuracies for $M=1$ and $M=2$, see Fig.~\ref{fig:tdMclock}. This is likely related to the small peak in the histograms at $\tau = 2\pi/\Omega_{\rm m}$, see Fig.~\ref{fig:M1_vs_M2}. While it limits the accuracy in Sec.~\ref{sec:M2}, it is almost completely absent in the additional mean-field approximation as can be seen in Fig.~\ref{fig:Mclock}. As $M$ increases, we expect this peak to vanish, as it becomes increasingly unlikely that no tick occurs during a resonance.

\section{Conclusions \& Outlook}
\label{sec:conclusions}

We introduced an optomechanical system as a model for an autonomous mechanical pendulum clock driven solely by thermal resources. Our work opens a new avenue to investigate the fundamental limitations of time-keeping by demonstrating that pendulum clocks can be operated in the quantum regime, driven by the radiation pressure of single photons that are emitted one-by-one into a cavity. This pendulum clock violates the TUR and is thus more accurate for a given entropy production than any clock described by classical rate equations or overdamped Langevin equations. By including multiple emitters within the optical cavity, we demonstrated the system's evolution towards deterministic, noiseless clock behavior expected from macroscopic classical pendulum clocks. In contrast to previous findings~\cite{Gopal_2024_electronicclock}, the TUR violations do not decrease as we go towards this regime, across the quantum-to-classical transition. We note that, despite the experimental accessibility of optomechanical systems, achieving the specific parameters required for ideal clock operation currently poses a significant challenge, mainly due to the large single-photon optomechanical coupling that is required.

An intriguing direction for future research lies in the exploration of correlations between the emitters. For instance, the phenomenon of superradiance~\cite{dicke:1954,kirton_2019} could result in a scaling of the accuracy in $M$ that is more favorable, resulting in larger TUR violations as the number of emitters is increased, similar to the findings in Ref.~\cite{meier:2024}. To this end, our theoretical model could be extended to allow for correlations between the emitters. Another avenue left for future work is to follow Ref.~\cite{Wadhia:2025} and investigate the entropy production associated to turning the microscopic ticks into a macroscopic signal used for filtering.

All the necessary codes and instructions to reproduce the
simulations in our work can be found in this repository~\cite{mohammad_mehboudi_2026_20026922}.
\begin{acknowledgments}
We acknowledge discussions with J. Arnold, C. Bruder, A. Roulet, S. Seah, F. Meier, M. Huber and R. Silva in the early stage of the project. M.B. and P.P.P. acknowledge funding from the Swiss National Science Foundation under Grant No. PCEFP2194268. This research was funded in part by the Austrian Science Fund (FWF) (``NOQUS,'' Grant DOI: 10.55776/PAT1969224, and ``Thermal Machines in the quantum world,'' Grant DOI: 10.55776/I6047) and by the European Research Council (Consolidator Grant No. ``Cocoquest'' 101043705). M.B. acknowledges funding from the European Research Council (ERC) under the European Union’s Horizon 2020 research and innovation program (Grant Agreement No. 101002955–CONQUER). N.B. acknowledges financial support from the Swiss National Science Foundations (NCCR-SwissMAP). The authors acknowledge TU Wien Bibliothek for financial support through its Open Access Funding Programme.
\end{acknowledgments}

\bibliography{library}
\appendix
\onecolumngrid

\section{Analytical steady-state expression in the limit of a rigid cavity}\label{app:three-level-laser}

In this Appendix, we provide additional information on the steady state of the optomechanical pendulum clock~\eqref{eq:FullME} in the absence of optomechanical coupling $g=0$. In this limit the mechanical degrees of freedom decouple  and the emitter-cavity system settles into a stationary steady state, which describes a three-level laser. For simplicity, we set $\bar n_\mathrm{f}=0$ in what follows.
First of all, we notice that in the absence of emitter-light coupling, i.e., for $f=0$, the emitter and the cavity factorize, the cavity decays to the vacuum while the emitter to a steady state given by
\begin{align}
\langle \hat p_{1}\rangle_{\mathrm{ss}}=\frac{\bar n_\mathrm{c}(1+\bar n_\mathrm{h})}{\bar n_\mathrm{c}+\bar n_\mathrm{h}+3\bar n_\mathrm{h}\bar n_\mathrm{c}}\,,\hspace{1cm}
\langle \hat p_{2}\rangle_{\mathrm{ss}}=\frac{\bar n_\mathrm{h}(1+\bar n_\mathrm{c})}{\bar n_\mathrm{c}+\bar n_\mathrm{h}+3\bar n_\mathrm{h}\bar n_\mathrm{c}}\,,\hspace{1cm}\langle \hat p_{3}\rangle_{\mathrm{ss}}=\frac{\bar n_\mathrm{h}\bar n_\mathrm{c}}{\bar n_\mathrm{c}+\bar n_\mathrm{h}+3\bar n_\mathrm{h}\bar n_\mathrm{c}}\,.
\end{align} 
From these expression, the population inversion $\langle \hat \sigma_z\rangle_\mathrm{ss}=\langle\hat{p}_2\rangle_{\mathrm{ss}}-\langle\hat{p}_1\rangle_{\mathrm{ss}}$ is easily computed and reads
\begin{equation}\label{eq:InvUncoupled}
\langle \hat \sigma_z\rangle_\mathrm{ss}=\frac{\bar n_\mathrm{h}-\bar  n_\mathrm{c}}{\bar n_\mathrm{c}+\bar n_\mathrm{h}+3\bar n_\mathrm{h}\bar n_\mathrm{c}} \,,
\end{equation}
from which is clear that population inversion is achieved as long as $\bar n_\mathrm{h}>\bar n_\mathrm{c}$.

In the presence of emitter-light coupling $f\neq 0$ (but still $g=0$), it is still possible to obtain an analytic solution for the steady state of Eq.~\eqref{eq:FullME}, provided that we factorize the third-order cumulant $\left\langle \hat{a}^\dagger\hat{a}(\hat{p}_2 - \hat{p}_1)\right\rangle \rightarrow \langle \hat a^\dagger \hat a\rangle (\langle \hat p_2\rangle - \langle \hat p_1\rangle)$. While explicit expressions are quite unwieldy, we can find insightful relations between stationary quantities, as for instance
\begin{equation}\label{eq:Inv}
\langle \hat \sigma_z\rangle_\mathrm{ss}=\frac{\bar n_\mathrm{h}-\bar n_\mathrm{c}}{\bar n_\mathrm{c}+\bar n_\mathrm{h}+3\bar n_\mathrm{h}\bar n_\mathrm{c}}
- \frac{\kappa  \bigl(\gamma_\mathrm{h}(2+3\bar n_\mathrm{h})+\gamma_\mathrm{c}(2+3\bar n_\mathrm{c})\bigr)}{\gamma_\mathrm{h} \gamma_\mathrm{c} (\bar n_\mathrm{c}+\bar n_\mathrm{h}+3\bar n_\mathrm{h}\bar n_\mathrm{c})}\langle \hat a^\dagger \hat a\rangle_\mathrm{ss} \,.
\end{equation}
The first term on the righthand side is the same as~\eqref{eq:InvUncoupled}, to which gets subtracted a non-negative term, i.e., the population inversion gets lowered in the presence of light-matter coupling, as it sustains lasing. 
We also find that the light-matter correlations are proportional to the number of photons
\begin{equation}
\langle \hat a^\dagger \hat a\rangle_\mathrm{ss}= \frac{2f}{\kappa} \mathrm{Im}\langle \hat a^\dagger \sigma_{12}\rangle_\mathrm{ss}\,,
\end{equation}
and  
\begin{equation}
 \mathrm{Re}\langle \hat a^\dagger \sigma_{12}\rangle_\mathrm{ss}=-\frac{2\Delta}{\kappa+\gamma_\mathrm{h}\bar n_\mathrm{h}+\gamma_\mathrm{c}\bar n_\mathrm{c}} \mathrm{Im}\langle \hat a^\dagger \sigma_{12}\rangle_\mathrm{ss}\,.
\end{equation}
Large values of the mean photon number can be reached even when population inversion has not yet occurred, in which case one speaks of lasing without inversion. 
An important working point to characterize is the inversion point. We can enforce $\langle \sigma_z\rangle_\mathrm{ss}=0$ and obtain an expression for the threshold value of coupling, i.e., to the strength needed to achieve population inversion. For $\Delta=0$ and $\bar n_\mathrm{h}>\bar n_\mathrm{c}$, we obtain 
\begin{equation}
f_\mathrm{th}=\frac{1}{2}\sqrt{\frac{(\bar n_\mathrm{h}-\bar n_\mathrm{c})\gamma_\mathrm{h} \gamma_\mathrm{c}(\kappa+\gamma_\mathrm{h}\bar n_\mathrm{h}+\gamma_\mathrm{c}\bar n_\mathrm{c})}{\gamma_\mathrm{c}(\bar n_\mathrm{c}+1)+\gamma_\mathrm{h}(\bar n_\mathrm{h}+1)}}\,.
\end{equation}

\section{Benchmarking the approximate stochastic master equation}
\label{app:benchmark}
\begin{figure}
    \centering
    \includegraphics[width=1\columnwidth]{ 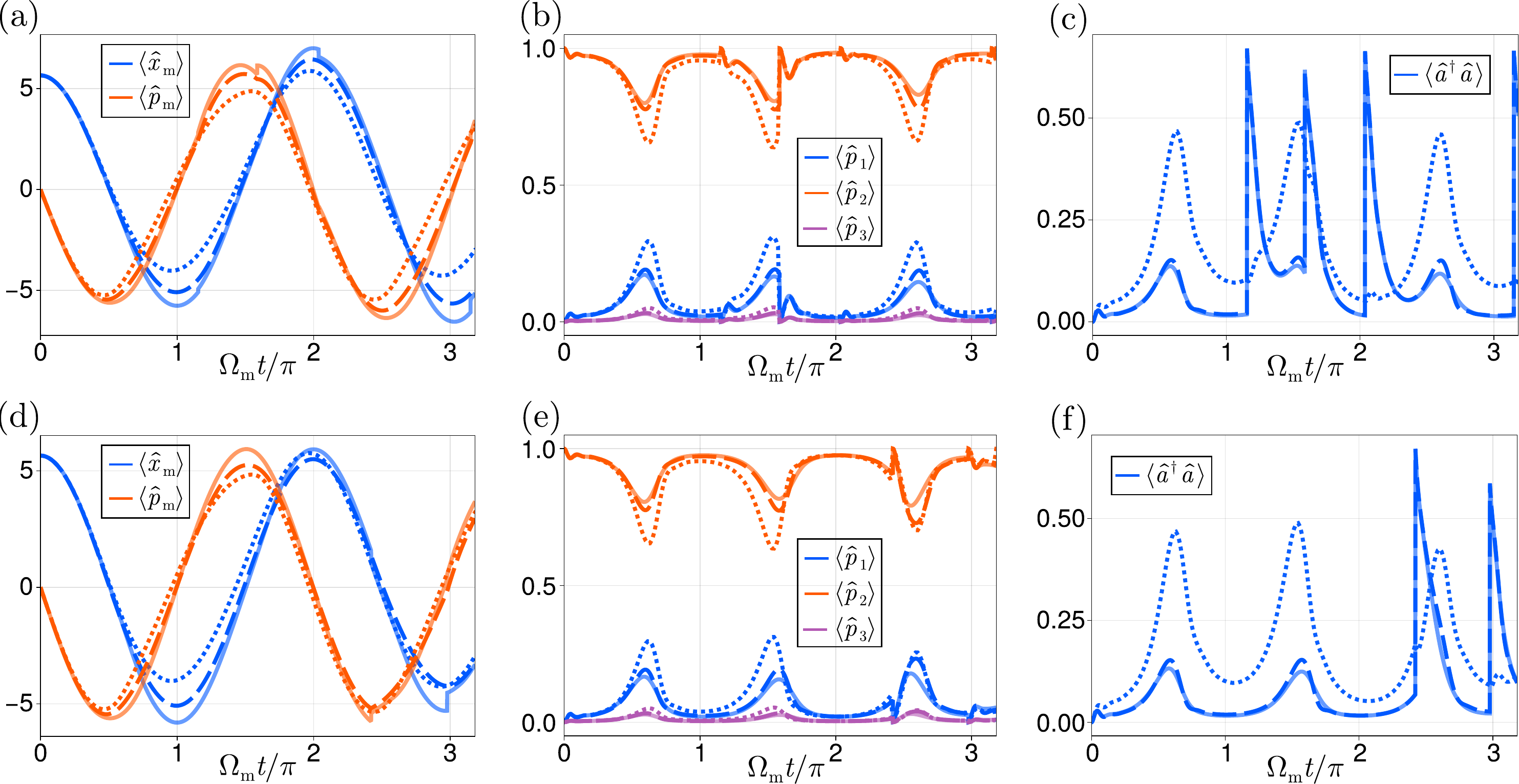}
    \caption{Comparing approximations to the quantum pendulum clock. Solid lines are obtained from simulating a trajectory of Eq.~\eqref{eq:fullunrav}. Dashed lines are obtained from Eq.~\eqref{eq:stochmas2}, together with Eqs.~\eqref{eq:xm} and \eqref{eq:pm}. Dotted lines are obtained from Eqs.~\eqref{eq:eqsMi}-\eqref{eq:q2} for $M=1$, together with Eqs.~\eqref{eq:xm} and \eqref{eq:pm}. (a) Mechanical quadratures. Correlations between the mechanical oscillator and the emitter influence the mechanical motion slightly. (b) Atom populations. The populations are well approximated by assuming the mechanical resonator to factorize. Assuming the emitter and the cavity to factorize results in the correct qualitative behavior. (c) Cavity occupation. The average number of photons in the cavity is well approximated by assuming the mechanical resonator to factorize but correlations between the cavity and the emitter have a strong impact. Parameters: $\omega_{13} = 240\gamma_{\rm h}, \omega_{23} = \Omega_{\rm f} = 120\gamma_{\rm h}$ ($\Delta=0$),$\Omega_{\rm m}=\gamma_{\rm h}$, $\gamma_{\rm c} = 50\gamma_{\rm h}$, $\gamma_{\rm m}=0.01\gamma_{\rm h}$, $\kappa = 3.4\gamma_{\rm h}$, $f=2.4\gamma_{\rm h}$, $g=2.6\gamma_{\rm h}$, $\bar{n}_{\rm h}=10$, (a)-(c) $\bar{n}_{\rm c}=\bar{n}_{\rm f}=\bar{n}_{\rm m}=0$, (d)-(f) $\bar{n}_{\rm c}=5\cdot 10^{-4}$ and $T_{\rm c}=T_{\rm f}=T_{\rm m}$.}
    \label{fig:benchmark}
\end{figure}

Here we benchmark Eq.~\eqref{eq:stochmas2} against an unraveled version of Eq.~\eqref{eq:FullME} for parameters that result in smaller mechanical amplitudes, where simulations are feasible. To this end, we unravel Eq.~\eqref{eq:FullME}, resulting in
\begin{equation}
    \label{eq:fullunrav}
    d\hat\rho_{\rm tot} =\mathcal{L}_{\rm tot}\hat\rho_{\rm tot} dt+(dN-dt\mathrm{Tr}\{\mathcal{J}\hat\rho_{\rm tot}\})\left(\frac{\mathcal{J}\hat\rho_{\rm tot}}{\mathrm{Tr}\{\mathcal{J}\hat\rho_{\rm tot}\}}-\hat\rho_{\rm tot}\right),
\end{equation}
where $\mathcal{J}$ is defined in Eq.~\eqref{eq:jump} and we introduced $\mathcal{L}_{\rm tot}$ by re-writing Eq.~\eqref{eq:FullME} as 
\begin{equation}
    \frac{d}{dt}\varrho_{\rm tot}(t) = \mathcal{L}_{\rm tot}\varrho_{\rm tot}(t).
\end{equation}
To compare Eqs.~\eqref{eq:fullunrav} with Eq.~\eqref{eq:stochmas2}, we first solved Eq.~\eqref{eq:fullunrav} for a single realization of $dN(t)$ (a single trajectory) by drawing a random number between $0$ and $1$ in each time-step and setting $dN=1$ if this number is below $p=dt{\rm Tr}\{\mathcal{J}\varrho_{\rm tot}\}$. We then used this particular realization of $dN(t)$ to solve Eq.~\eqref{eq:jump}, together with Eqs.~\eqref{eq:xm} and \eqref{eq:pm} that describe the mechanical resonator. This ensures that we can compare solutions where the ticks happen at the exact same times. As illustrated in Fig.~\ref{fig:benchmark}, the approximate equations used in the main text describe the pendulum clock very well, in particular the cavity occupation and the emitter populations, which govern the behavior of the ticks. In Fig.~\ref{fig:benchmark}, we also include the solution of Eqs.~\eqref{eq:eqsMi}-\eqref{eq:q2} for $M=1$, together with Eqs.~\eqref{eq:xm} and \eqref{eq:pm}. They show that for a single emitter ($M=1$), correlations between the emitter and the cavity can become relevant, in particular in computing the cavity occupation.

These results indicate that assuming the mechanical resonator to factorize from the other constituents is indeed a good approximation, even for finite cold temperatures. For larger mechanical amplitudes, as they are present for the parameters used in the main text, we expect this approximation to become even better as the fluctuations in the mechanical resonator become smaller compared to the amplitude.

\section{Correlations between ticks}
\label{app:correlations}
\begin{figure}[ht]
    \centering
    \begin{overpic}[width=0.47\textwidth]{ 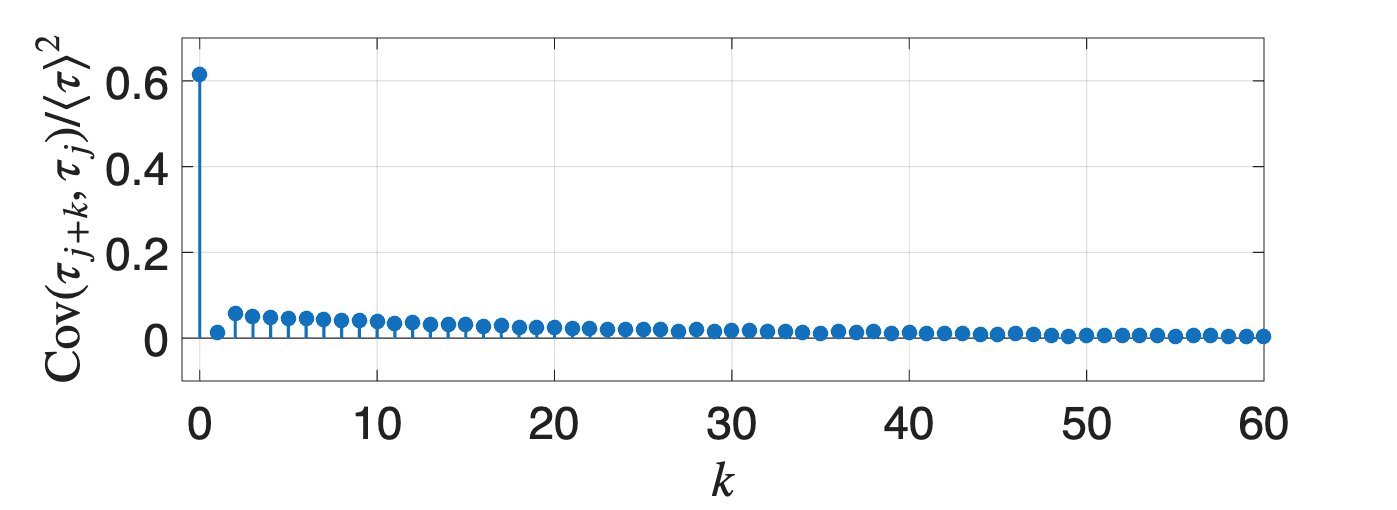}
        \put(16, 30){\large{(a)}}
    \end{overpic}
    \begin{overpic}[width=0.47\textwidth]{ 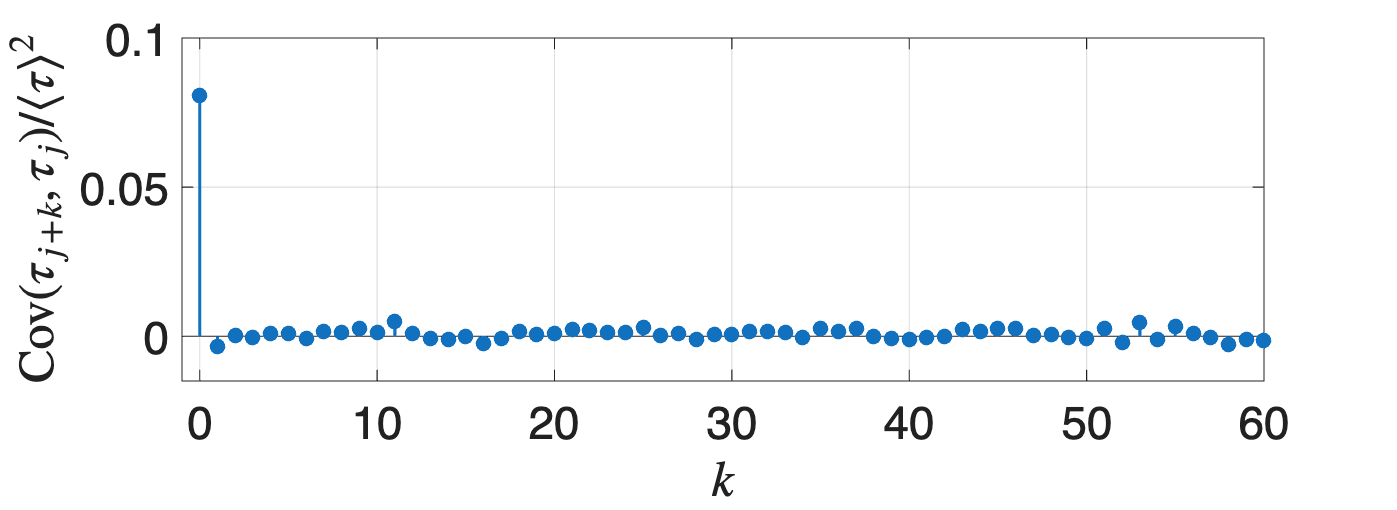}
        \put(16, 30){\large{(b)}}
    \end{overpic}
    \caption{ Covariance between a tick and the $k$-th tick thereafter. (a) unfiltered data. (b) filtered data. In both cases, correlations between subsequent waiting times are small. The parameters are the same as in Fig.~\ref{fig:histograms}.}
    \label{fig:acf}
\end{figure}
The performance of a clock is generally affected by correlations between consecutive ticks. To investigate these correlations, we consider the covariance between a tick and the $k-$th tick thereafter
\begin{equation}
    {\rm cov}(\tau_j,\tau_{j+k}) = \langle \tau_j\tau_{j+k}\rangle-\langle \tau\rangle^2,
    \label{eq:covariance}
\end{equation}
where $\tau_j = t_{j+1}-t_{j}$ denotes the $j$-th waiting time in a time-series of ticks occuring at $t_j$. These covariances are plotted in Fig.~\ref{fig:acf}. Both for the filtered and the unfiltered ticks, we find that the correlations between subsequent waiting times are small. For the unfiltered ticks, we find a minimum in Eq.~\eqref{eq:covariance} for $k=1$, i.e., waiting times following each other. We attribute this to the bimodal nature of the unfiltered waiting-time distribution. With multiple ticks occuring in a single resonance, a long waiting time (between resonances) is more likely followed by a short waiting time (within the same resonance).
For the filtered ticks, subsequent waiting times are weakly anti-correlated. This may be a consequence of to the asymmetry in the limit cycle, which results in two different waiting times during one mechanical period, see Figs.~\ref{f:PopCavEvo} and \ref{fig:trajectories}.  Our findings are in contrast to recent findings~\cite{culhane:2024}, where correlations between consecutive ticks have been found to decay slowly with the distance between the waiting times. This is likely due to the stochastic nature of the photo-emission process which we use to model the ticks.

\section{Computing heat and entropy production}
\label{app:thermodynamics}

Due to the approximative nature of our master equation, we can only compute approximations of heat currents. While it is possible to ensure that a Markovian master equation ensures the laws of thermodynamics~\cite{potts:2021}, this is not the case for our master equation here. Nevertheless, the approximations taken are well justified and we do not observe any violations of the laws of thermodynamics for the parameter ranges that we consider. 

\begin{figure}
    \centering
    \includegraphics[width=0.70\linewidth]{ 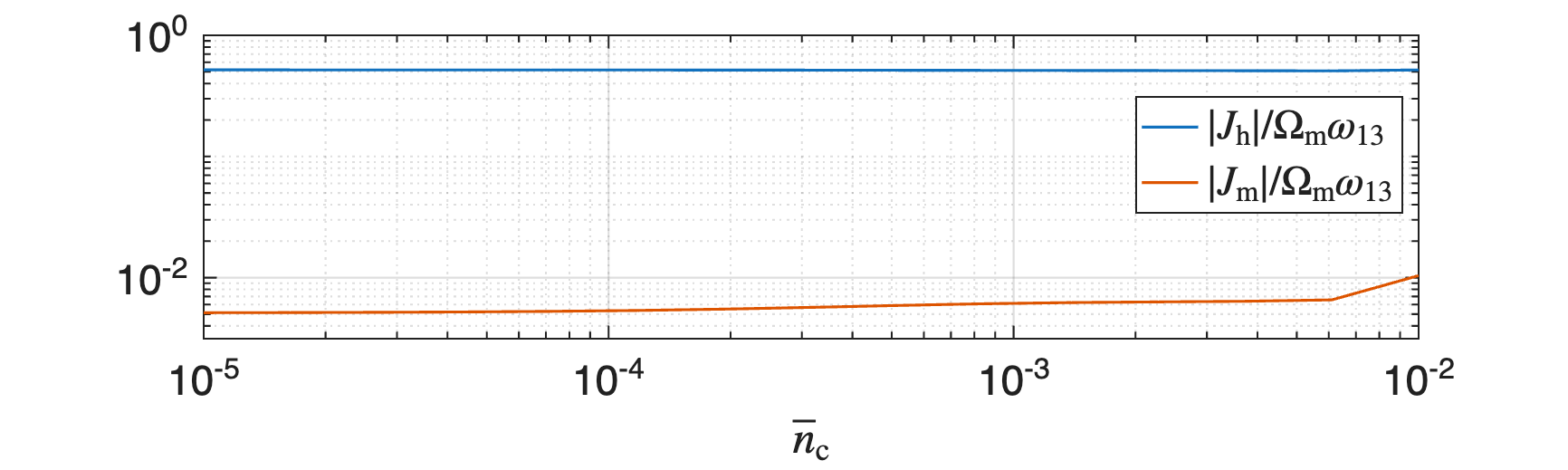}
    \caption{The heat currents from the hot bath, and from the mechanical bath shows that one can ignore the latter in comparison. Parameters are the same as in Fig.~\ref{f:LimitCycle}.}
    \label{fig:heat_vs_heat}
\end{figure}

To shed further insight into the dissipation of our quantum pendulum clock, we note that we have a separation of energy scales (see Sec.~\ref{sec:ideal-clock})
\begin{equation}
    \label{eq:sepens}
    \omega_{13},\,\omega_{23},\,\omega_{12},\, \Omega_{\rm f} \gg \Omega_{\rm m},\,\kappa,\,\gamma_\alpha,\,f,\,g.
\end{equation}
Per tick, the heat exchanged with the hot (cold) bath is of the order of $\omega_{13}$ ($\omega_{23}$). Similarly, the heat exchanged with the bath of the cavity is of the order of $\Omega_{\rm f}$. Since the phonon number exchanged with the mechanical oscillator is of the order of one per tick, the heat exchanged with the mechanical bath is of the order of $\Omega_{\rm m}$ and thus negligible compared to the other contributions to the heat current. This is illustrated in Fig.~\ref{fig:heat_vs_heat}. For our pendulum clock, the vast majority of dissipation is therefore responsible for producing the ticks, not for maintaining the periodic motion of the pendulum. This is in contrast to macroscopic pendulum clocks, where the force of friction plays a much bigger role than in nanoscale optomechanical devices.

Indeed, our formalism does not allow for computing the heat dissipated by the mechanical oscillator. The reason is that the energy exchanged in the optomechanical coupling changes the frequency of the photons~\cite{Aspelmeyer_2014_review}. Within the Markovian master equation, we cannot keep track of the frequency at which the photons leave the resonator and we therefore do not have access to the energy that the photons lose to the mechanics. Interesting avenues for a more detailed investigation into the dissipation of our pendulum clock are provided by Refs.~\cite{potts:2021,monsel:2021}.

Within the approximation that the dissipation of the mechanics is negligible, we can formally prove the laws of thermodynamics. To this end, we compute the inner energy using the free Hamiltonian in Eq.~\eqref{eq:hamfree}
\begin{equation}
    \label{eq:firstlawapp1}
    \partial_t \langle \hat{H}_0\rangle = -i\langle [\hat{H}_0,\hat{H}]\rangle + J_\mathrm{f}+ J_\mathrm{h}+ J_\mathrm{c}+ J_\mathrm{m},
\end{equation}
where the heat currents are given by
\begin{equation}
    J_\alpha = -{\rm Tr}\{\hat{H}_0\mathcal{L}_\alpha\hat\varrho_{\rm tot}\},
\end{equation}
in agreement with Eq.~\eqref{eq:heathot}. We note that for $\Delta=0$, we find $[\hat{H}_0,\hat{H}] = \Omega_\mathrm{m}[\hat{b}^\dagger\hat{b},\hat{H}]$ and thus
\begin{equation}\label{eq:J_m_ss_1}
    \Omega_\mathrm{m}\partial_t\langle \hat{b}^\dagger\hat{b} \rangle = -i\langle [\hat{H}_0,\hat{H}]\rangle + J_\mathrm{m}.
\end{equation}
In the steady state, we thus find
\begin{equation}
    \label{eq:firstlawapp2}
 J_\mathrm{f}+J_\mathrm{h}+J_\mathrm{c} = 0,
\end{equation}
which is the same as the first law stated in the main text, Eq.~\eqref{eq:firstlaw}, upon neglecting $J_\mathrm{m}$. By using Eq.~\eqref{eq:J_m_ss_1} and setting $\partial_t\langle {\hat b}^{\dagger} \hat b \rangle = 0$, one can straightforwardly calculate the heat current to the mechanical bath as
\begin{align}
    J_m = -\sqrt{2}g \langle {\hat a}^{\dagger} \hat a {\hat p}_m\rangle \approx -\sqrt{2}g \langle {\hat a}^{\dagger} \hat a \rangle \langle {\hat p}_m\rangle,
\end{align}
which as depicted in Fig.~\ref{fig:heat_vs_heat}, is negligible compared to the heat exchanged with the hot bath.

The second law of thermodynamics can be derived by casting the heat currents into the form
\begin{equation}
    \frac{J_\alpha}{T_\alpha} = k_B{\rm Tr}\{(\mathcal{L}_\alpha\hat\varrho)\ln \hat\varrho_\alpha\},
\end{equation}
where the state
\begin{equation}
    \hat\varrho_\alpha = \frac{e^{-\beta_\alpha\hat{H}_0}}{{\rm Tr}\{e^{-\beta_\alpha\hat{H}_0}\}},
\end{equation}
with $\beta_\alpha = 1/k_BT_\alpha$, is a fixed point of the corresponding dissipator, i.e., $\mathcal{L}_\alpha\varrho_\alpha=0$.
By further writing the derivative of the von Neumann entropy as
\begin{equation}
    \label{eq:vonneu}
    \partial_t S_\mathrm{vN}[\hat\varrho_{\rm tot}] = -\partial_t{\rm Tr}\{\hat\varrho_{\rm tot}\ln\hat\varrho_{\rm tot}\} = -\sum_{\alpha=\rm f,h,c,m}{\rm Tr}\{(\mathcal{L}_\alpha\hat\varrho_{\rm tot})\ln \hat\varrho_{\rm tot}\},
\end{equation}
we can cast the entropy production into
\begin{equation}
    \dot\Sigma = k_B\partial_t S_\mathrm{vN}[\hat\varrho_{\rm tot}] + \sum_{\alpha=\rm f,h,c,m}\frac{J_\alpha}{T_\alpha} = -\sum_{\alpha=\rm f,h,c,m}{\rm Tr}\{(\mathcal{L}_\alpha\hat\varrho)(\ln \hat\varrho_{\rm tot}-\ln \hat\varrho_\alpha\}\geq 0,
\end{equation}
where the last inequality is Spohn's inequality~\cite{spohn:1978,spohn:1978b}. In the steady state, this directly reduces to the second law of thermodynamics stated in the main text, Eq.~\eqref{eq:2ndlaw}.

\section{The quantum pendulum clock with many emitters}
\label{app:Mnoise}
Here we consider the quantum pendulum clock with many emitters, and we show that in the limit $M\rightarrow \infty$, the clock as well as the readout of its ticks becomes noiseless. To this end, we consider Eqs.~\eqref{eq:eqsMi}-\eqref{eq:q2}. We first note that one may be tempted to rescale $f\rightarrow f/M$, in order to remove the $M$-dependence in Eq.~\eqref{eq:eqsMi}. This would however suppress the buildup of coherence between the emitters and the photons, see Eq.~\eqref{eq:asM}, and the clock would no longer work. Physically, reducing $f$ suppresses spontaneous emission into the cavity. Since the cavity empties during each cycle of the clock, spontaneous emission is crucial for filling the cavity with photons each time the mechanical oscillator is in resonance with the cavity.

To investigate the large $M$ regime, we assume that there exists a time interval $d\tau$ during which many jumps occur, but the averages involving the emitters stay approximately constant. This is justified since the emitter operators $\hat{q}_j$ and $\hat{\sigma}$ are averages over all emitters, see Eq.~\eqref{eq:atomops}, and thus only slightly affected by a single jump.
To make progress, we follow App.~B of Ref.~\cite{landi:2024}. Within the time-interval $d\tau$, the jumps are uncorrelated, since the probability conditioned on all previous jumps given in Eq.~\eqref{eq:probM} does not change. Therefore, the time-integral of $dN$ is well described by a binomial distribution, i.e.
\begin{equation}
    P\left(\int_t^{t+d\tau}dN(t')\right) = B[d\tau/dt,\langle \hat{q}_3\rangle \gamma_\mathrm{c}(\bar n_\mathrm{c} + 1)dt],
\end{equation}
where $B[n,p]$ is the Binomial distribution with $n$ attempts and probability $p$. For large $n$, the Binomial distribution is well approximated by a Gaussian with average $np$ and variance $np(1-p)$. With this, one may show that the quantity
\begin{equation}
    dW = \frac{1}{\sqrt{M\langle \hat{q}_3\rangle \gamma_\mathrm{c}(\bar n_\mathrm{c} + 1)}}\int_t^{t+d\tau}dN(t) - d\tau \sqrt{M\langle \hat{q}_3\rangle \gamma_\mathrm{c}(\bar n_\mathrm{c} + 1)},
\end{equation}
is a Wiener increment, i.e., a Gaussian random variable with zero mean and variance of $d\tau$. With the help of this Wiener increment, we can integrate the equations of motion (only of the averages involving the emitter) and we find

\begin{equation}
\label{eq:appasM}
\begin{aligned}
    d\left\langle \hat{a}^\dagger \hat{\sigma} \right\rangle  =  &d\tau\left[i\Delta  -\frac{1}{2}(\kappa+\gamma_{\rm h}\bar n_{\rm h} + \gamma_{\rm c} \bar n_{\rm c})\right]\left\langle \hat{a}^\dagger \hat{\sigma} \right\rangle -i\sqrt{2}g d\tau\left\langle \hat{a}^\dagger \hat{\sigma}\right\rangle \langle\hat{x}_\mathrm {m}\rangle  + ifd\tau\left[\langle \hat{q}_2\rangle+ \overline{\left\langle \hat{a}^\dagger\hat{a}\right\rangle} (\langle\hat{q}_2\rangle - \langle\hat{q}_1\rangle) \right]\\ 
& -  \left\langle \hat{a}^\dagger \hat{\sigma} \right\rangle\sqrt{\frac{\langle \hat{q}_3\rangle \gamma_\mathrm{c}(\bar n_\mathrm{c} + 1)}{M}}dW,
\end{aligned}
\end{equation}
\begin{equation}
\label{eq:appq1}
     d\langle \hat{q}_1\rangle   =2fd\tau {\rm Im} \left\langle \hat{a}^\dagger \hat{\sigma} \right\rangle + \gamma_{\rm h} d\tau(\bar n_{\rm h} +1)\langle \hat{q}_3\rangle - \gamma_{\rm h} d\tau\bar n_{\rm h} \langle \hat{q}_1\rangle 
    - \langle \hat{q}_1\rangle\sqrt{\frac{\langle \hat{q}_3\rangle \gamma_\mathrm{c}(\bar n_\mathrm{c} + 1)}{M}}dW, 
\end{equation}
\begin{equation}
\label{eq:appq2}
    d\langle \hat{q}_2\rangle  = -2f d\tau{\rm Im} \left\langle \hat{a}^\dagger \hat{\sigma} \right\rangle + \gamma_{\rm c}d\tau(\bar n_{\rm c} +1)\langle \hat{q}_3\rangle - \gamma_{\rm c}d\tau\bar n_{\rm c} \langle \hat{q}_2\rangle 
    +  (1-\langle \hat{q}_2\rangle)\sqrt{\frac{\langle \hat{q}_3\rangle \gamma_\mathrm{c}(\bar n_\mathrm{c} + 1)}{M}}dW,
\end{equation}
where we introduced the averaged photon number
\begin{equation}
    \overline{\left\langle \hat{a}^\dagger\hat{a}\right\rangle} = \frac{1}{d\tau}\int_t^{t+d\tau} dt'\left\langle \hat{a}^\dagger\hat{a}\right\rangle.
\end{equation}
Note that the photon number, as well as the mechanical resonator, are still determined by the stochastic differential equations given in Eqs.~\eqref{eq:eqsMi}, \eqref{eq:xm}, and \eqref{eq:pm}, with time increment $dt$. These averages may vary considerably over the time-scale $d\tau$, since each jump results in an additional photon in the cavity (in the limit $\bar{n}_\mathrm{c}=0$), and the limit cycle of the mechanics grows with $M$ as a consequence.

According to Eqs.~\eqref{eq:appasM}, \eqref{eq:appq1}, and \eqref{eq:appq2}, the noise becomes less and less relevant as $M$ increases and we obtain deterministic averages in the limit $M\rightarrow \infty$. While the mean photon number and the averages of the mechanical oscillator may vary during the time-scale $d\tau$, they also become deterministic for $M\rightarrow \infty$ since they are only affected by the noise indirectly, through the emitter.
We may also introduce a diffusive current
\begin{equation}
    \label{eq:diffcurr}
    I_{\rm diff}(t)d\tau = \frac{1}{M}\int_t^{t+d\tau} d N(t') = \langle \hat{q}_3\rangle \gamma_\mathrm{c}(\bar n_\mathrm{c} + 1)d\tau + \sqrt{\frac{\langle \hat{q}_3\rangle \gamma_\mathrm{c}(\bar n_\mathrm{c} + 1)}{M}}dW,
\end{equation}
which may be computed from the observation of the unfiltered ticks. In the large $M$ limit, this measurable current directly provides the average occupation $\langle \hat{q}_3\rangle$, from which perfect tick statistics may be obtained since $\langle \hat{q}_3\rangle$ becomes deterministic in the large $M$ regime.

From Eq.~\eqref{eq:diffcurr}, we may anticipate the scaling of the accuracy in the limit of large $M$. To this end, we expect the accuracy of any clock constructed from the diffusive current to be limited by its noise
\begin{equation}
    \label{Idiffacc}
    \mathcal{N}\propto \frac{\langle I_{\mathrm{diff}}\rangle}{\langle\!\langle I_{\mathrm{diff}}^2\rangle\!\rangle},
\end{equation}
with
\begin{equation}
    \label{Idiffnoise}
    \langle\!\langle I_{\mathrm{diff}}^2\rangle\!\rangle = \lim_{T\rightarrow\infty}\frac{1}{T}\int_0^Tdt\int_{-\infty}^\infty ds\left\{E[I_\mathrm{diff}(t+s)I_\mathrm{diff}(t)]-E[I_\mathrm{diff}(t+s)]E[I_\mathrm{diff}(t)]\right\},
\end{equation}
where the expectation value $E[\bullet]$ denotes the average over all trajectories. To motivate Eq.~\eqref{Idiffacc}, we note that for a renewal process, $\mathcal{N}=\langle \tau\rangle^2/\langle\!\langle\tau^2\rangle\!\rangle = {\langle I\rangle}/{\langle\!\langle I^2\rangle\!\rangle}$~\cite{landi:2024}. We now assume that the stochasticity in $I_\mathrm{diff}$ is dominated by the last term in Eq.~\eqref{eq:diffcurr}, i.e., we treat $\langle \hat{q}_3\rangle$ as a deterministic quantity obeying $E[\langle\hat{q}_3\rangle X]=\langle\hat{q}_3\rangle E[X]$ for any quantity $X$. We then find
\begin{equation}
    \label{eq:diffnoiseacc}
    \langle\!\langle I_{\mathrm{diff}}^2\rangle\!\rangle = \frac{\langle \hat{q}_3\rangle \gamma_\mathrm{c}(\bar n_\mathrm{c} + 1)}{M}\hspace{.75cm}\Rightarrow\hspace{.75cm}\mathcal{N} \propto M.
\end{equation}

\end{document}